\documentclass[twocolumn,floatfix,pra,aps,showpacs,amsmath,superscriptaddress]{revtex4-1}
\usepackage{graphicx,float,tabularx,epstopdf,color,dcolumn,bm,amsfonts,amsmath}

\DeclareMathOperator{\Tr}{Tr}

\begin{document}
\title{Dynamics in multiple-well Bose-Einstein condensates }

\author{M. Nigro}
\author{P. Capuzzi} 
\author{H. M. Cataldo} \author{D. M. Jezek}
\affiliation{Universidad de Buenos Aires, Facultad de Ciencias Exactas
  y Naturales,  Departamento de F\'{i}sica, Buenos Aires, Argentina}
\affiliation{IFIBA, CONICET-UBA, Pabell\'on 1, Ciudad Universitaria, 1428
  Buenos Aires, Argentina }
\begin{abstract}

  We study the dynamics of three-dimensional weakly linked
  Bose-Einstein condensates using a multimode model with an effective
  interaction parameter. The system is confined by a ring-shaped
  four-well trapping potential.  By constructing a two-mode
  Hamiltonian in a reduced highly symmetric phase space, we examine
  the periodic orbits and calculate their time periods both in the
  self-trapping and Josephson regimes.  The dynamics in the vicinity
  of the reduced phase space is investigated by means of a Floquet
  multiplier analysis, finding regions of different linear stability
  and analyzing their implications on the exact dynamics.  The
  numerical exploration in an extended region of the phase space
  demonstrates that two-mode tools can also be useful for performing a
  partition of the space in different regimes.  Comparisons with
  Gross-Pitaevskii simulations confirm these findings and emphasize
  the importance of properly determining the effective on-site
  interaction parameter governing the multimode dynamics.

\end{abstract}
\date{\today}
\maketitle
\section{Introduction}
The experimental research on Bose-Einstein condensates trapped in
ring-shaped optical lattices constitutes a promising area that is
meant to opening the possibility to study a rich emerging physics. In
consequence, important efforts are being made towards the effective
realization of such configurations \cite{amico1,hen09,jen16}.  For
instance, a lattice of tunnel junctions on a ring would enable the
creation of lattice models with periodic boundary conditions and with
the resulting ability to support piercing magnetic fluxes and the
associated topological phenomena \cite{jen16}.  In particular, it
would provide an ideal environment for the study of the Kibble-Zurek
mechanism, where the buildup of winding number in the phase transition
from Mott insulator to superfluid driven by tunneling rate increase,
is expected to occur, except for very slow quench times
\cite{zurek}. On the other hand, quantum information applications of
such configurations have begun to be devised, such as the
experimentally feasible qubit system based on bosonic cold atoms
trapped in ring-shaped optical lattices proposed by Amico {\it et al.}
\cite{amico}.  A practical implementation of this system could lead to
substantially lower decoherence rates, as the use of neutral atoms as
flux carriers would minimize the well-known characteristic
fluctuations in the magnetic fields of solid state Josephson qubits.

Concerning theoretical studies on this issue, the dynamics on ring
lattices with three \cite{trespozos2011} and four wells
\cite{cuatropozos06} have been previously investigated through
multimode (MM) models that utilized {\em ad-hoc} values for the
hopping and on-site energy parameters.  Substantial improvements were
reported in Ref. \cite{jezek13b}, where such parameters were
calculated {\em ab initio} by constructing a set of two-dimensional
localized wave functions in terms of the stationary solutions of the
Gross-Pitaevskii (GP) equation for a ring with an arbitrary number of
wells. This can be regarded as a similar procedure of the two-mode
(TM) model of a double-well condensate
\cite{smerzi97,ragh99,anan06,jia08,albiez05,mele11,abad11,doublewell},
where the order parameter is described as a superposition of wave
functions localized in each well with time dependent coefficients
\cite{smerzi97,ragh99}.  Such localized wave functions are
straightforwardly obtained in terms of the stationary symmetric and
antisymmetric states, which in turn determine the parameters involved
in the TM equations of motion \cite{smerzi97,ragh99,anan06,jia08}. The
corresponding dynamics exhibits Josephson and self-trapping regimes
\cite{smerzi97,ragh99} which have been experimentally observed by
Albiez {\it et al.}  \cite{albiez05}.  The self-trapping (ST)
phenomenon, which is also present in extended optical lattices
\cite{optlat,Anker2005,Wang2006}, is a non linear effect where the
difference of populations between neighboring sites does not change
sign during the whole time evolution.  There is nowadays an active
research on the ST effect, which involves different types of systems,
including mixtures of atomic species \cite{stlastoplat,mele11}.

In recent works it has been shown that the on-site interaction energy
dependence on the population imbalance has to be taken into account
for the TM model, in order to accurately describe the exact dynamics in
double-well systems \cite{jezek13a,jezek13b,nosEPJD}. Such an
imbalance dependence gives rise to a reduced effective on-site
interaction energy parameter when it is introduced into the equations
of motion of the model. In the Thomas-Fermi approximation it has been
shown that such a parameter is reduced by a factor of 7/10, 3/4, or
5/6 depending on the dimensionality of the system. Later, it has been
proven that the effective on-site interaction energy parameter is also
fundamental to describe the dynamics in a ring-shaped lattice, within
the frame of MM models in two-dimensional condensates as well
\cite{jezek13b}.

The phase space of a MM dynamics in a $N_c$-well system has $2N_c-2$
dimensions.  Hence, the analysis of such a dynamics for $N_c\ge 3$
does not pose a simple task to handle. The goal of this work is to
show that useful results can still be obtained by using mathematical
tools such as symmetry criteria and special techniques developed for
non-linear differential equations \cite{Floquet}.  In particular, we
will numerically treat a three-dimensional four-site ring-shaped
optical lattice.  The construction of its multimode parameters will be
based on previous works \cite{jezek13b,cat11}, where a method to
obtain localized on-site Wannier-like (WL) functions in a ring-shaped
optical lattice was developed. These states are obtained as a
superposition of stationary states of the GP equation with different
winding numbers.  Here we will show how to optimally localize these WL
states to finally obtain the effective on-site interaction energy
parameter for furnishing an accurate model.  On the other hand, by
restricting the dynamics to a symmetric case, we will construct a
two-mode type Hamiltonian able to predict transitions to the ST
regime. With such a Hamiltonian we will calculate the orbit periods in
both Josephson and ST regimes.  Next, we will show that the location
of the two-mode critical point of the Josephson to ST transition turns
out to be quite useful to determine the domains of different regimes
in an extended region compared to that of the symmetric case.
These findings will be confirmed by local TM models involving
only pairs of neighboring sites \cite{Anker2005,Wang2006}.
Finally, by calculating Floquet multipliers \cite{Floquet} we
will analyze the dynamical stability in the surroundings of the
TM solutions and we will show that in the more stable regions a
criteria for calculating characteristic times can be established.

This paper is organized as follows.  In Sec. \ref{multi} we describe
the trapping potential and include the equations of motion of the MM
model. Next, we explain the procedure for obtaining the localized
states used to describe the dynamics and analyze the conditions to
achieve maximally localized WL functions. To conclude this section, we
summarize the method for calculating the effective on-site energy
parameter and analyze the corresponding results of a few
representative configurations.  In Sec. \ref{sec:sym} we numerically
study the dynamics, showing that one can predict the ST and Josephson
regimes using a reduced-space Hamiltonian that describes high-symmetry
systems. This reduction allows us to extend previous analytic results
of the period of the trajectories in the TM model to these systems.
Next in Sec.\ \ref{sec:floquet}, we study the dynamics close to this
highly symmetric situation by means of a Floquet analysis. Finally in
Sec. \ref{sec:regN1N3}, we numerically obtain the MM dynamics for
non-symmetric configurations in the vicinity of the symmetric
condition, establishing useful connections to the symmetric case
results and comparing with several full GP solutions.  To conclude, a
summary of our work is presented in Sec. \ref{sum}.  The definition of
the parameters employed in the equations of motion are gathered in the
Appendix \ref{sec:parameters}, while in Appendix \ref{sec:appB} we
give some details on the Floquet analysis theory.

\section{ The multimode model}\label{multi}

 \subsection{ The trap  }\label{trap4}

\begin{figure}
\begin{center}
\includegraphics[width=0.85\columnwidth,clip=true]{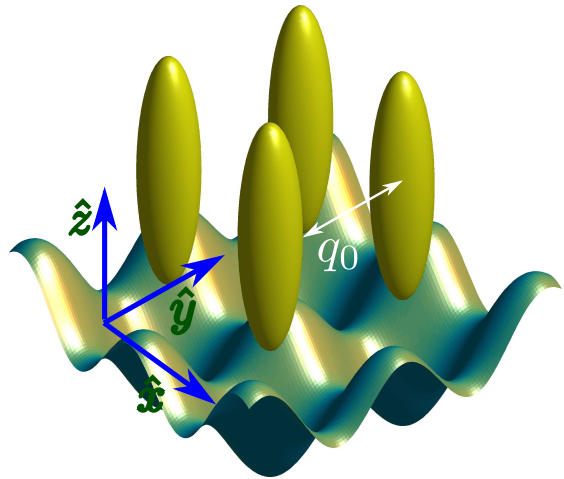}
\end{center}
\caption{\label{fig:rho3D} (color online) Isosurfaces of the
  ground-state density and the trapping potential of the four-site
  system with $N=10^4$ and $V_b/\hbar\omega_x=25$ (in arbitrary
  units).}
\end{figure}

 We consider a three-dimensional Bose-Einstein condensate of Rubidium
 atoms confined by the external trap
\begin{align}
V_{\text{trap}}({\bf r} ) =& \frac{ 1 }{2 } \, m \, \left[
\omega_{x}^2  x^2  + \omega_{y}^2  y^2 
+ \omega_{z}^2  z^2 \right] \nonumber \\
&+   
V_b \left[ \, \cos^2(\pi x/q_0)+   
\, \cos^2(\pi y/q_0)\right],
\label{eq:trap4}
\end{align}
where $m$ is the atom mass. The harmonic frequencies are given by
$ \omega_{x}= \omega_y= 2 \pi \times 70 $ Hz, and
$ \omega_{z}= 2 \pi \times 90 $ Hz, and the lattice parameter is
$ q_0= 5.1 \mu$m.  The barrier height parameter $V_b $ and the number
of particles will take different values upon the calculation. For
instance, in Fig. \ref{fig:rho3D} we have plotted isosurfaces of the
ground-state density and the trapping potential for $N= 10^4 $ and
$V_b/(\hbar\omega_x)=25$.  Hereafter, time and energy will be given in
units of $\omega_x^{-1}$ and $\hbar\omega_x$, respectively.

\subsection{Equations of motion}

For completeness, in this section we will sketch the procedure for
obtaining the equations of motions reported in Ref. \cite{jezek13b}.
The multimode order parameter for $N_c$ sites is expressed in terms of
$N_c$ localized WL functions $w_k$ as,
\begin{equation}
\psi ({\mathbf r})_{MM} = \sum_{k} \,  b_k(t)  \,  w_k ({ r, \theta,z})
  \,.
\label{orderparameter}
\end{equation}
where $-[(N_c-1)/2]\le k \le [N_c/2]$.  Inserting the above expression
into the time dependent GP equation, the equations of motion for the
coefficients $b_k(t)= e^{i \phi_k} \, |b_k| $ are obtained, which can
be cast into $ 2 N_c $ real equations for the populations
$ n_k =|b_{k}|^2 = N_k / N $ and the phase differences
$ \varphi_k= \phi_k - \phi_{k-1}$ of each site as,
\begin{widetext}
\begin{align}
 \hbar\,\frac{dn_k}{dt} =&  - 2 J [ \sqrt{n_k \, n_{k+1}} \, \sin\varphi_{k+1} 
-\sqrt{n_k \, n_{k-1} } \, \sin\varphi_k ]\nonumber\\
&-  2 F [ \sqrt{n_k \, n_{k+1} } (n_k + n_{k+1} ) \, \sin\varphi_{k+1}
-\sqrt{n_k \, n_{k-1} } (n_k + n_{k-1} ) \, \sin\varphi_k]
\label{ncmode1hn}
\end{align}
\begin{align}
 \hbar\,\frac{d\varphi_k}{dt}  = & N (U_{k-1}n_{k-1} - U_kn_k) 
-  J \left[ \left(\sqrt{\frac{n_k}{ n_{k-1}}} - \sqrt{\frac{n_{k-1} }{ n_k}}\,\right) \, \cos\varphi_k
+ \sqrt{\frac{n_{k-2}}{ n_{k-1} }} \, \cos\varphi_{k-1} 
- \sqrt{\frac{n_{k+1} }{ n_k}} \, \cos\varphi_{k+1}\right]\nonumber\\
&-  F \left[ \left( n_k \sqrt{\frac{n_k}{ n_{k-1} }} - n_{k-1} \sqrt{\frac{n_{k-1}}{ n_k}}\,\right)
 \, \cos\varphi_k 
+ \left(3\, \sqrt{n_{k-2} \, n_{k-1}} + n_{k-2} \sqrt{\frac{n_{k-2}}{ n_{k-1}}}\,\right) 
 \, \cos\varphi_{k-1} \right.\nonumber\\
&- \left.\left(3\, \sqrt{n_{k+1} \, n_k} + n_{k+1} \sqrt{\frac{n_{k+1}} { n_k}}\,\right)  \, 
\cos\varphi_{k+1}\right],
\label{ncmode2hn}
\end{align}
\end{widetext}
where $U_k$ is the on-site interaction energy in the $k$-site. The
bare MM model assumes a constant $U_k=U$ value.  In contrast, in the
effective MM model its dependence on the imbalance is considered
$U_k=U_k(\Delta N_k)$, which gives rise to a reduced effective
parameter $U_{\text{eff}}$ (see Sec.\ \ref{sec:Ue}). The definition of
the bare interaction parameter $U$ and the tunneling parameters $J$
and $F$ are given in the Appendix \ref{sec:parameters}.  As the
populations and phase differences must fulfill $\sum_k n_k=1$ and
$\sum_k \varphi_k=0$, respectively, only $2N_c- 2$ equations are
independent. In Eq.\ (\ref{ncmode2hn}) we have excluded the terms
involving the overlap between the localized densities in each site, as
this parameter turns out to be two orders of magnitude smaller than
the rest of the tunneling parameters, $J$ and $F$.

\subsection{ Localized states and multimode model parameters}

In this section we summarize the procedure to obtain the localized
states \cite{cat11,jezek13b} necessary to describe the dynamics. The
choice of these states is not unique, as they inherits the freedom in
the choice of the global phase of stationary states. In the field of
atomic and molecular physics it has been long applied the concept of
localized molecular orbitals, or ``Boys orbitals'' in chemistry, and
after that in electronic calculations in periodic systems (see e.g the
review of Ref.\ \cite{mar12} and references therein) in order to
optimize the basis set used.  We will therefore analyze how the
localization of the WL functions affect the determination of the model
parameters, especially the on-site energy parameter $U$.

The stationary states $\psi_n( r, \theta, z )$ are obtained as the
numerical solutions of the three-dimensional GP equation \cite{gros61}
with different winding numbers $n$ \cite{je11,jezek13b}.  Assuming
large barrier heights \cite{cat11}, the winding numbers will be
restricted to the values $-[(N_c-1)/2]\leq n \leq [N_c/2]$
\cite{je11}.  We have shown in Ref. \cite{cat11} that
stationary states of different winding number are orthogonal, and can
be used to define localized, orthogonal, WL functions on each $k$
site. These are given by
\begin{equation}
w_k({ r, \theta, z })=\frac{1}{\sqrt{N_c}} \sum_{n} \psi_n({ r, \theta, z})
 \, e^{-i n\theta_k } \,,
\label{wannier}
\end{equation}
where $\theta_k=2\pi k/N_c$.

The ground state ($n=0$) and the state with maximum winding number,
$n=2$ for the four-site system, have completely uniform phases in each
well \cite{je11}. Both functions can be chosen to be real with
$\psi_0 >0 $ and $\psi_2 >0$ in the first quadrant
($0<\theta<\pi/2$). This means that we have fixed their phases to zero
in that quadrant.  On the other hand, the winding numbers with
$n=\pm 1 $ correspond to vortex-like states, which have an associated
velocity field that gives rise to a non vanishing angular momentum
\cite{je11}.  Although their velocity fields are very small in each
site, the phase is not absolutely uniform and to perform the sum of
Eq. (\ref{wannier}) it is important to correctly choose the global
phases of $\psi_1$ and $\psi_{-1}$ to obtain maximum localization.  In
our case, without loosing generality one can set the phases of
$ \psi_1({ r, \theta, z}) $ and $\psi_{-1}(r, \theta, z)$ to zero at
the bisectrix $\theta= \pi/4$ and $z=0$. And taking into account that
$\psi_1=\psi_{-1}^*$ it is sufficient to consider a single variational
parameter $\eta$ in their phases as $e^{\pm i \eta}\psi_{\pm 1}$ to
analyze the localization of the Wannier function.  The localized state
$w_k$ as a function of $ \eta $ thus acquires the form,
\begin{align}
w_k({ r, \theta,z , \eta})=&\frac{1}{2} \left[
\psi_0({ r, \theta, z}) + \psi_2({ r, \theta, z})\cos k\pi \right.
\nonumber \\ 
&\left. +  2 \mathrm{Re}( e^{i (\eta-k\pi/2)} \psi_1({ r, \theta,z})) \right]   \,,
\label{wannier0}
\end{align}
with the conditions on each $\psi_n({ r, \theta, z}) $ given above.

\begin{figure}
\includegraphics[width=\columnwidth,clip=true]{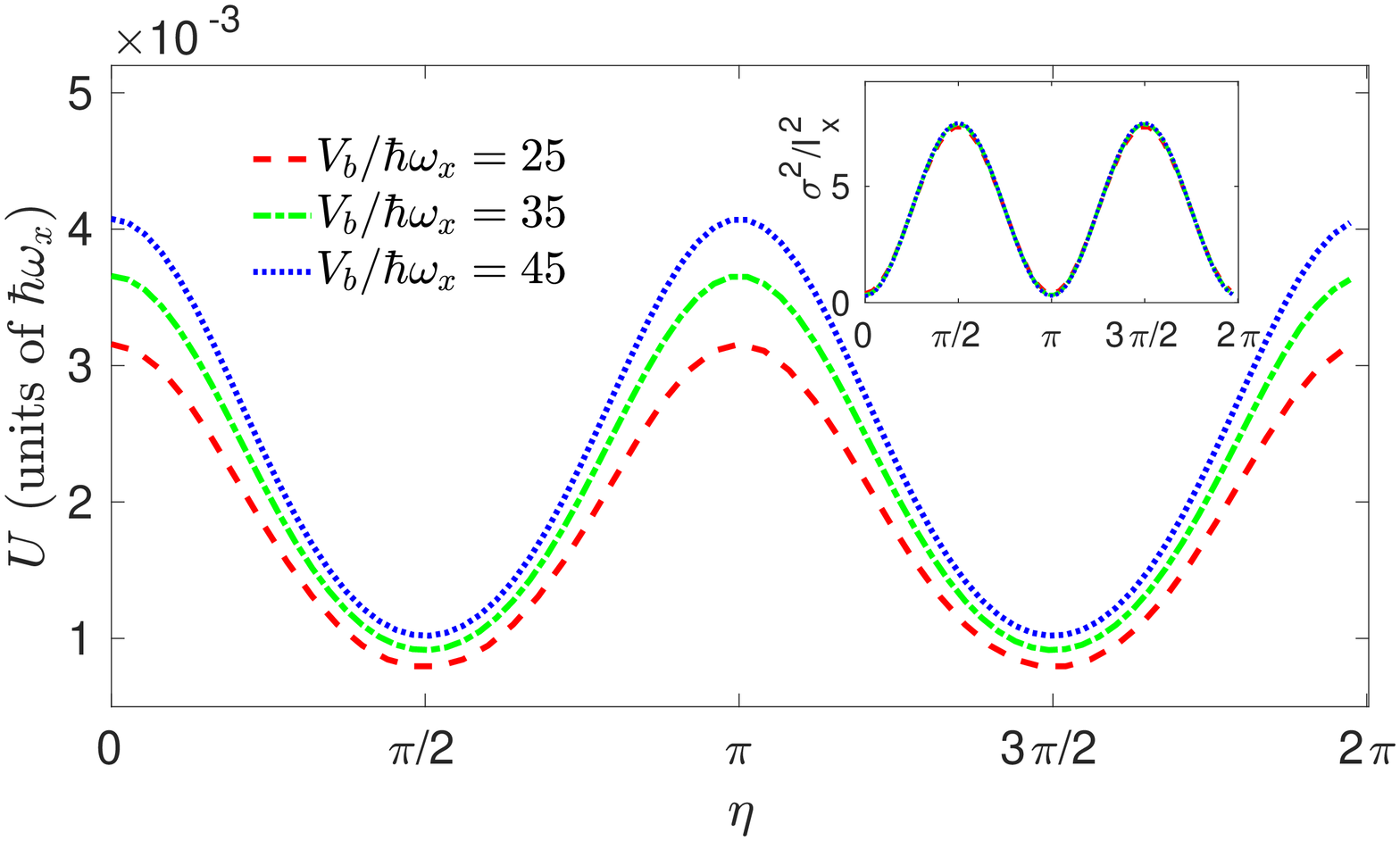}
\caption{\label{fig:Us_eta} Bare on-site interaction energy parameter
  $U$ as a function of the phase $\eta$. The inset shows the
  dispersion $\sigma$ of the WL functions in the $xy$ plane as a
  function of $\eta$, with $l_x=l_y=\sqrt{\hbar/(m\omega_x)}$}
\end{figure}

The maximum localization is achieved by minimizing the spatial
dispersion of the WL wave functions in the $xy$ plane
$ \sigma^2= \langle x^2 + y^2\rangle -(\langle x\rangle ^2 + \langle
y\rangle^2)$
with respect to $\eta$.  The degree of localization of the WL wave
functions strongly affects the values of the model parameters. This is
shown in Fig.\ \ref{fig:Us_eta} where we depict the calculated on-site
interaction energy $U$ as a function of $\eta$, together with the
dispersion $\sigma$. Whereas the hopping parameters turn out to be
rather independent of this phase, the parameter $U(\eta)$ has shown to
be strongly dependent. Such a variation would qualitatively alter the
dynamics predicted by multimode models, thus demonstrating the
importance of a properly localized wave function.

\begin{figure}
\vspace*{1.5em}
\includegraphics[width=0.95\columnwidth,clip=true]{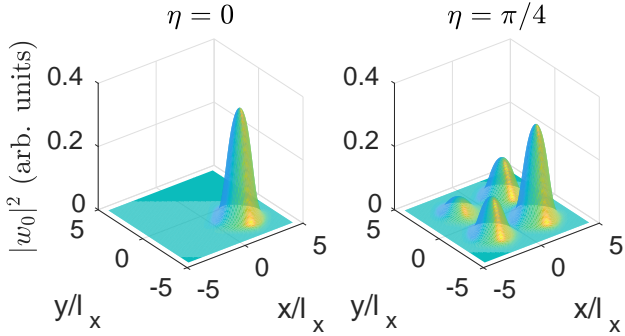}
\caption{\label{fig:w0wpi4} WL wave function densities $|w_0|^2$ (in
  arbitrary units) at the $z=0$ plane for $\eta=0$ (left panel) and
  $\eta=\pi/4$ (right panel).}
\end{figure}

In Fig. \ref{fig:w0wpi4} we show the three dimensional WL function
density $ w_0^2$ at $z=0$ with $\eta = 0 $ and $\eta = \pi/4 $,
showing that in the first case it is clearly more localized.

In summary, to obtain an accurate MM dynamics one should achieve the
maximum localization of the WL functions, which is found to be
fulfilled when the phases of all stationary states are chosen equal at
the bisectrix of a given site.

\subsection{\label{sec:Ue}Inclusion of effective on-site interaction effects}
In addition to the localization effects, to construct an accurate
model we need to calculate the effective on-site interaction energy
parameter $U_{\text{eff}}$. For that matter, we follow the procedure
described in Ref.\ \cite{jezek13b} valid for a ring-shaped lattice
with equal wells.  We thus first numerically calculate the on-site
interaction energy $U_k(\Delta N_k)$ in the $k$ site as a function of
$\Delta N_k =N_k - N/N_c$ as \cite{jezek13b}
\begin{equation}
  \frac{U_k(\Delta N_k)}{U} = \frac{\int d^3\mathbf{r}\rho_N(\mathbf{r})\rho_{N+\Delta N}(\mathbf{r})}{\int d^3\mathbf{r}\rho_N^2(\mathbf{r})},
\label{eq:Uk}
\end{equation}
with the normalized-to-unity ground-state densities $\rho_N$ and
$\rho_{N+\Delta N}$ for four-well systems of $N$ and $N+\Delta N$ total
number of particles, respectively, where $\Delta N = N_c\Delta N_k$.
The on-site interaction energy $U_{k}$ exhibits a linear dependence on
$\Delta N_k $ and it can be approximated by
\begin{equation}
\frac{U_{k}(\Delta N_k)}{U}  \simeq  1  -   \alpha   \frac{ N_c  \Delta N_k} {N} .
\label{UekN} 
\end{equation}

Replacing $ U_{k-1}(\Delta N_{k-1})$ and $ U_{k}(\Delta N_k)$ given by
the above equation in Eq. (\ref{ncmode2hn}), the effective multimode
(EMM) model equations of motion read 
%
%
%
%
\begin{widetext}
\begin{align}
 \hbar\,\frac{dn_k}{dt} = & - 2 J [ \sqrt{n_k \, n_{k+1}} \, \sin\varphi_{k+1} 
-\sqrt{n_k \, n_{k-1} } \, \sin\varphi_k ]\nonumber\\
&-  2 F [ \sqrt{n_k \, n_{k+1} } (n_k + n_{k+1} ) \, \sin\varphi_{k+1}
-\sqrt{n_k \, n_{k-1} } (n_k + n_{k-1} ) \, \sin\varphi_k]
\label{encmode1hn}
\end{align}
\begin{align}
 \hbar\,\frac{d\varphi_k}{dt}  = &  f_{3D}  ( n_{k-1} -n_{k}) N  U  -  
 \alpha (  n_{k-1} - n_{k}) N U [   N_c (n_{k-1}+  n_{k})-2 ] \nonumber\\
&-  J \left[ \left(\sqrt{\frac{n_k}{ n_{k-1}}} - \sqrt{\frac{n_{k-1} }{ n_k}}\,\right) \, \cos\varphi_k
+ \sqrt{\frac{n_{k-2}}{ n_{k-1} }} \, \cos\varphi_{k-1} 
- \sqrt{\frac{n_{k+1} }{ n_k}} \, \cos\varphi_{k+1}\right]\nonumber\\
&-  F \left[ \left( n_k \sqrt{\frac{n_k}{ n_{k-1} }} - n_{k-1} \sqrt{\frac{n_{k-1}}{ n_k}}\,\right)
 \, \cos\varphi_k 
+ \left(3\, \sqrt{n_{k-2} \, n_{k-1}} + n_{k-2} \sqrt{\frac{n_{k-2}}{ n_{k-1}}}\,\right) 
 \, \cos\varphi_{k-1} \right.\nonumber\\
&- \left.\left(3\, \sqrt{n_{k+1} \, n_k} + n_{k+1} \sqrt{\frac{n_{k+1}} { n_k}}\,\right)  \, 
\cos\varphi_{k+1}\right],
\label{encmode2hn}
\end{align}
where $f_{3D}= 1-\alpha$, and hence we obtain
$U_{\text{eff}}=f_{3D}U$.
\end{widetext}

In Table \ref{tab:2} we quote the values of $\alpha$ and $f_{3D}$ for
a few configurations, and observe that for larger particle numbers and
higher barrier heights, the parameter $f_{3D}$ approaches from above
the three-dimensional Thomas-Fermi limiting value of $ 7/ 10 $,
derived for the double-well model \cite{jezek13a}.  Also, it is
worthwhile noticing that the second term of the rhs of Eq.\
(\ref{encmode2hn}) is a second order correction on the population
imbalance, which in general does not give rise to noticeable changes in the
dynamics.

\begin{table}
  \caption{\label{tab:2} Linear coefficient $\alpha$ of the 
    on-site interaction  energy  $U_k$  and factor  $f_{3D}$ for  $ N_c =4 $  
    and different number of particles and barrier heights.}
\begin{ruledtabular}
\begin{tabular}{lcccc}
$ N $ & $   V_b/(\hbar\omega_x)  $ & $   \alpha $ &   $ f_{3D} $    \\[3pt]
\hline \\[-5pt]
$10^3$  &  $  10  $ &  $  0.20 $ &   $  0.80  $   \\[3pt]
$10^3$  &  $  15  $ &  $  0.20 $ &   $  0.80  $  \\[3pt]
$10^4$  &  $  25 $ &  $  0.28 $ &    $   0.72  $  \\[3pt]
$10^4$  &  $  35 $ &  $  0.29 $ &    $   0.71  $  \\[3pt]
\end{tabular}
\end{ruledtabular}
\label{table}
\end{table}

\section{\label{sec:dyn}The dynamics}
\subsection{ \label{sec:sym}TM symmetric case }

The orbits of the dynamical equations lie in a six-dimensional space,
and thus it becomes challenging to classify them taking into account
the possible different features.  In particular, it is important to
predict the regions of self-trapped and Josephson trajectories, where
in the multi-well system we define a self-trapped site $k$ as a site
whose population difference with neighboring sites $N_k-N_{k\pm 1}$
does not change sign during the whole time evolution
\cite{cuatropozos06}.  A subset of such trajectories can be found by
restricting the dynamics to a more symmetric case, which can be
described with a TM Hamiltonian.

Therefore, we will first analyze the multimode model in a symmetric
case where $n_k = n_{k+2}$ and $ \varphi_k= \varphi_{k+2}$.  In this
case, the second term on the right hand side of Eq.~(\ref{encmode2hn})
vanishes, and the only difference between the MM and EMM equations of
motion is given by the use of the effective interaction parameter
$ U_{\mathrm{eff}}$ instead of the bare $ U $.

Defining the imbalance $ Z = 2 (N_0 - N_1)/N $, the phase
difference $ \varphi= \varphi_1$, and $K= 2J+ F $, the equations of
motion Eqs. (\ref{encmode1hn}) and (\ref{encmode2hn}) reduce to
\begin{equation}
 \hbar \frac{dZ}{dt} = - 2 K \sqrt{1-Z^2}\sin\varphi    \, ,
\label{imbe}
\end{equation}
\begin{equation}
  \hbar \frac{d\varphi}{dt} = \frac{ U_{\mathrm{eff}}}{2} N  Z + 2 K \frac{ Z}{\sqrt{ 1 - Z^2}} \cos\varphi  \,.
\label{phasee}
\end{equation}
Then, changing for convenience the time units to $\hbar/2K$, one can
obtain a TM-type Hamiltonian for the reduced space,
\begin{equation}
 H ({Z},\varphi) = \frac{1}{2} \Lambda {Z}^2 -   \sqrt{1-{Z}^2}\cos\varphi ,
\label{eq:Hred}
\end{equation}
where $ \Lambda = {U_{\mathrm{eff}} N}/{(4 K)}$. 

We can thus obtain the critical imbalance between the Josephson and ST
regimes,
\begin{equation}
 {Z}_c = 
2 \frac{\sqrt{\Lambda-1 }}{\Lambda} ,
\label{eq:Zct}
\end{equation}
and calculate the exact time periods $T^{\mathrm{EMM}}$ using
$\Lambda$ \cite{ragh99,nosEPJD}, together with the approximations obtained in
the small-oscillation limit
\begin{equation}
T_{\mathrm{so}}^{\mathrm{EMM}}= \frac{\pi\hbar}{ K \sqrt{\Lambda + 1}}  \,,
\label{eq:tpeqoscr}
\end{equation}
and in the ST regime \cite{nosEPJD}
\begin{equation}
  T_{\mathrm{st}}^{\mathrm{EMM}}= \frac{ {Z}_i \pi\hbar}{ 2 K }\left( 1 - \sqrt{1- \frac{4}{\Lambda {Z}_i^2}}\right)  \, ,
\label{tstbuenar}
\end{equation}
where ${Z}_i$ is the initial imbalance.  

\begin{figure}
\includegraphics[width=0.9\columnwidth,clip=true]{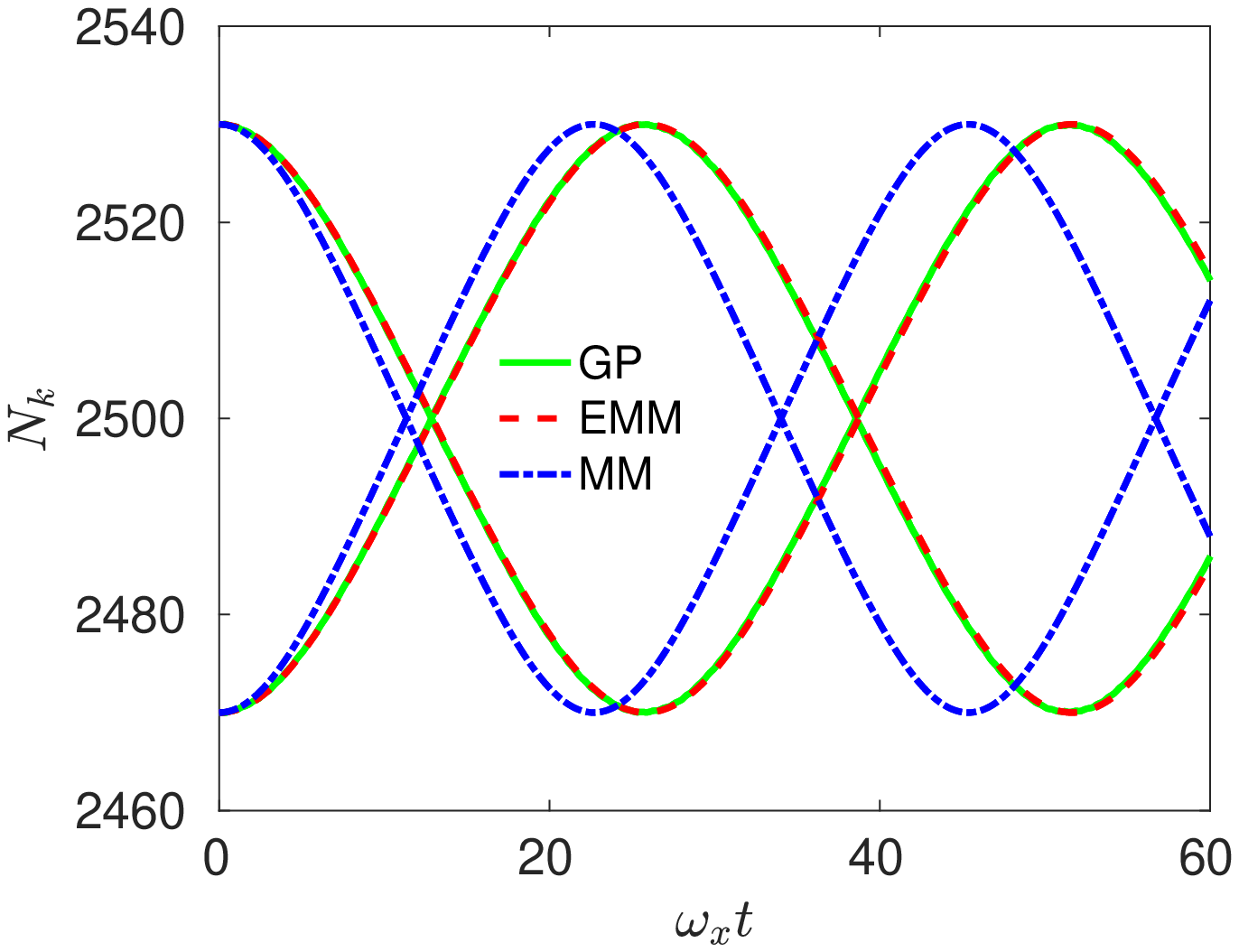}
\includegraphics[width=0.9\columnwidth,clip=true]{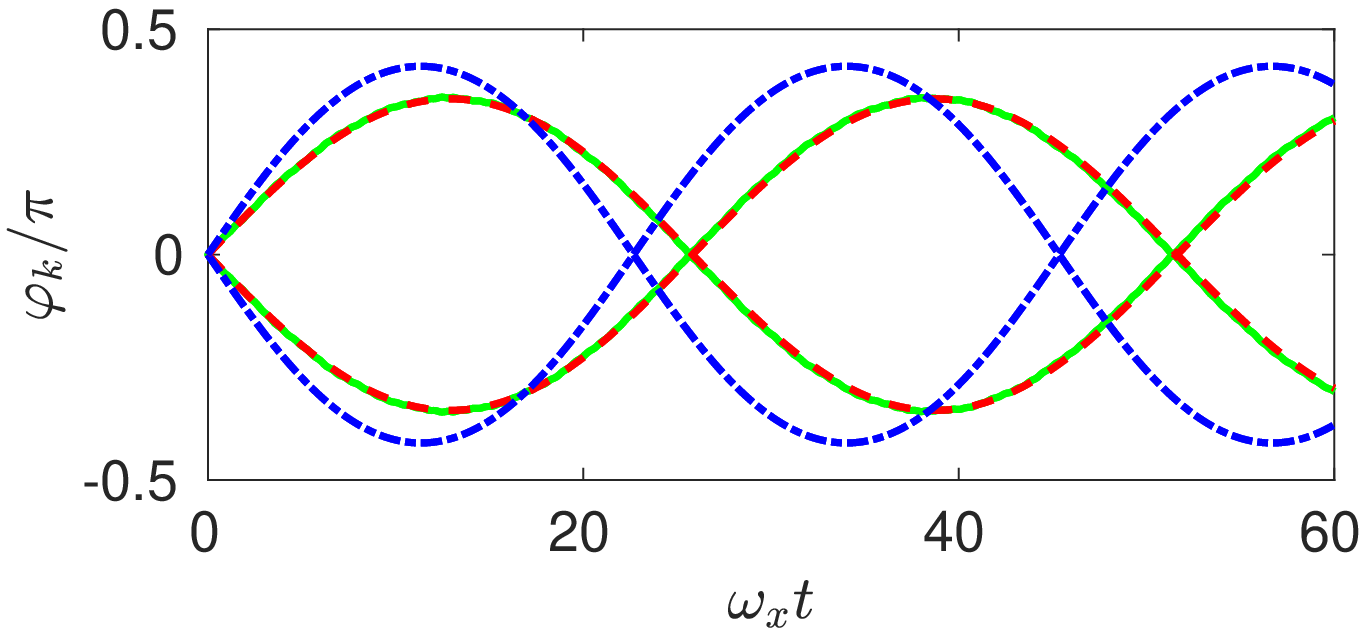}
\caption{\label{fig:joseph4}(color online) Condensate dynamics
  arising from the GP equation, and from the EMM and the MM models for
  an initial condition $N_0=2530$ and $N_1=2470$ corresponding to the
  Josephson regime. Top panel: populations in each well $N_k$, bottom
  panel: phase differences $\varphi_k$. }
\end{figure}

For the system with $ N= 10^4 $ particles and a barrier height
$V_b = 25 $ $ \hbar \omega_x $, we have obtained
$ J=-6.60\times 10^{-4}\hbar\omega_x $,
$F= 2.08\times 10^{-3} \hbar\omega_x$,
$U= 3.16\times 10^{-3} \hbar\omega_x$, and
$U_{\mathrm{eff}}=2.27\times10^{-3}\hbar\omega_x$ which yields a
critical imbalance $ N {Z}_c = 232$ within the EMM model.  On the other
hand, using the bare value of $U$, we would have obtained a smaller
threshold $ N {Z}_c = 196 $.  For the same system the
small-oscillation period yields
$ T_{\mathrm{so}}^{\mathrm{EMM}}= 47.84 \omega_x^{-1} $, which may be
compared to that obtained by means of GP simulations,
$T_{\mathrm{so}}^{\mathrm{GP}}\simeq 47.7 \omega_x^{-1}$. In Fig.
\ref {fig:joseph4} we show the time evolution for the initial
$ N {Z}_i=120 < N {Z}_c $ in the Josephson regime, within the GP, MM,
and EMM frameworks. It is worthwhile noticing that although the period
for this value of $Z_i$ departs from the small-oscillation limit, the
GP and exact EMM are in good agreement.

\begin{figure}
\includegraphics[width=0.9\columnwidth,clip=true]{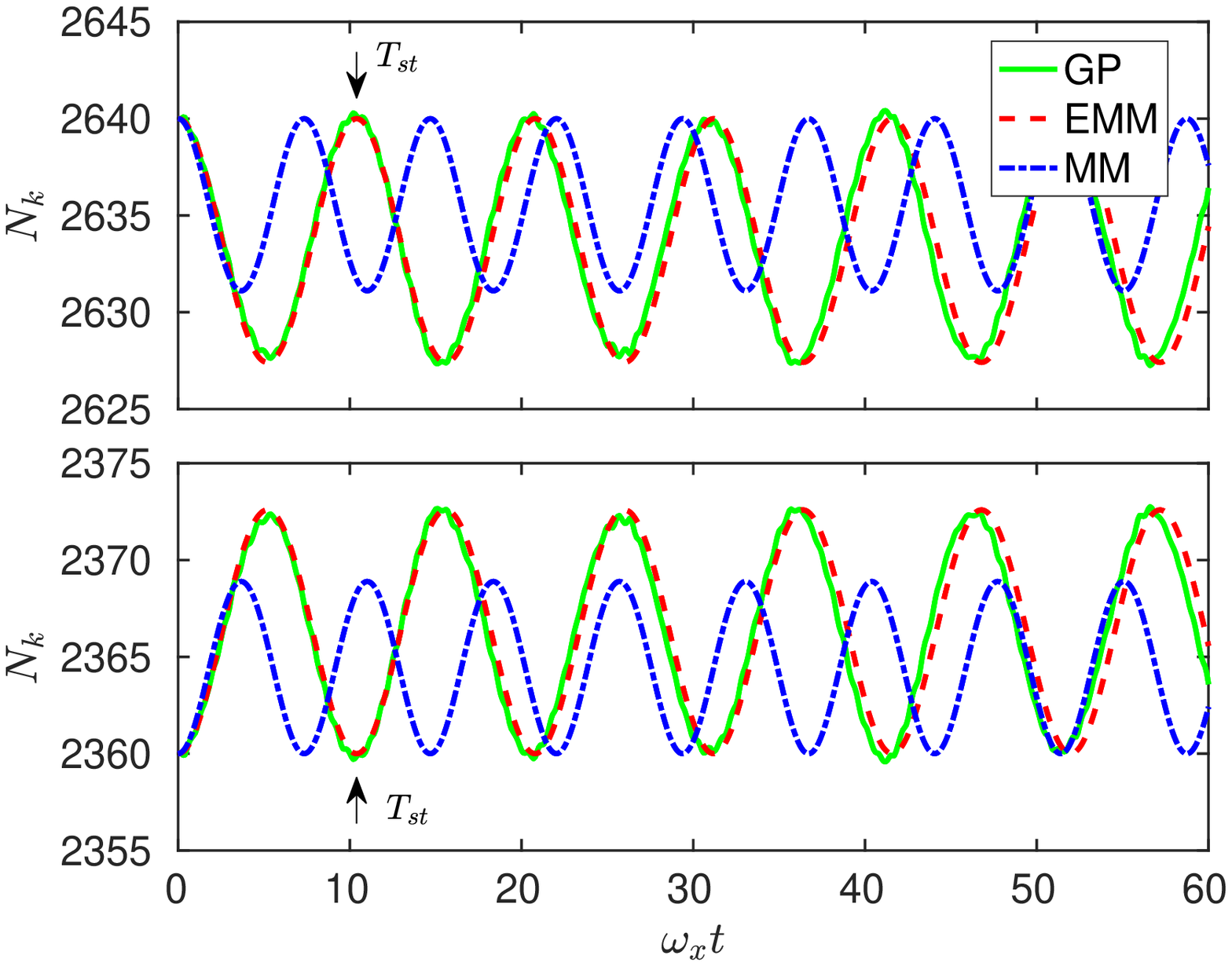}
\includegraphics[width=0.9\columnwidth,clip=true]{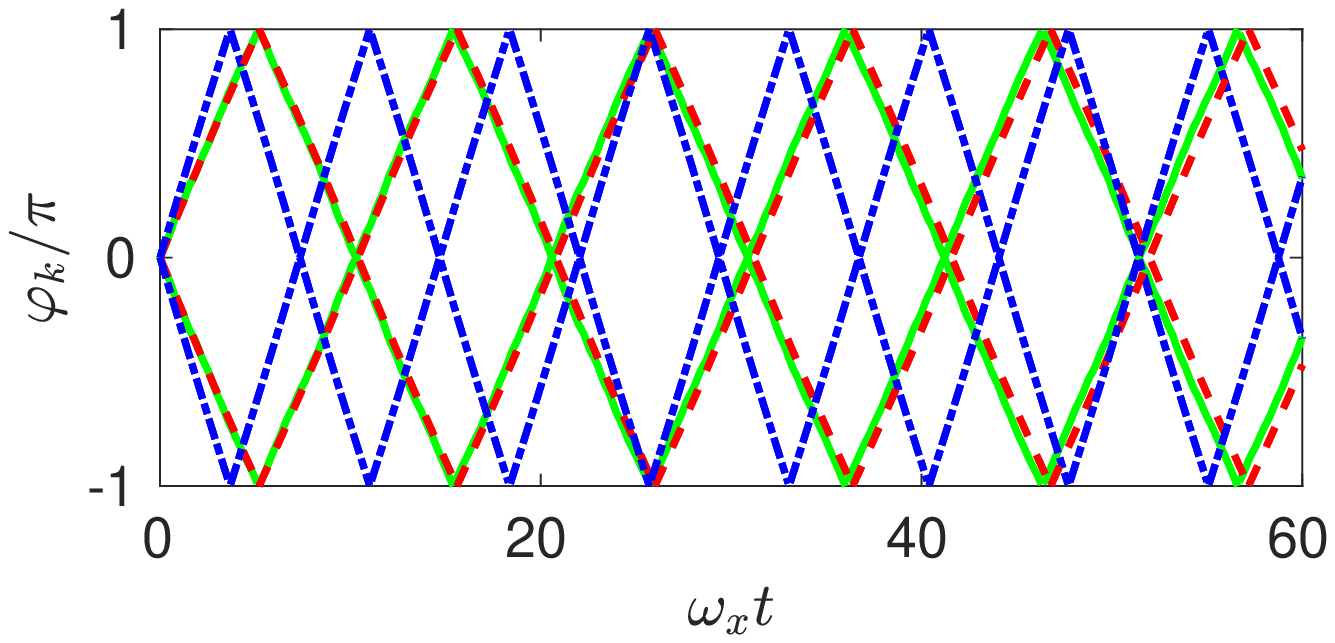}
\caption{\label{fig:self4}(color online) Condensate dynamics for the
  initial values $N_0=2640$, and $N_1=2360$ in the ST regime. We
  depict the GP simulation result, together with those arising from
  the EMM and MM models. Top panel: populations in each well $N_k$,
  bottom panel: phase differences $ \varphi_k$. }
\end{figure}

In Fig. \ref{fig:self4} we show the evolution of the populations in
each site for the initial condition $ N_0 = 2640$ and $N_1=2360$.  In
this case $ N {Z}_i = 560 > N{Z}_c $ and then the system
is in the ST regime. We may further calculate the period from Eq.\
(\ref{tstbuenar}), using in this case $ {Z}_i= 0.056 $ yielding
$ T_{\mathrm{st}}^{\mathrm{EMM}}= 10.391\, \omega_x^{-1} $, which turns out to
be in a good agreement with the GP simulation and EMM model results,
$T^{\mathrm{EMM}}=10.396\,\omega_x^{-1}$. 

\subsection{\label{sec:floquet}Near the TM symmetric case}
\begin{figure}
\includegraphics[width=0.9\columnwidth]{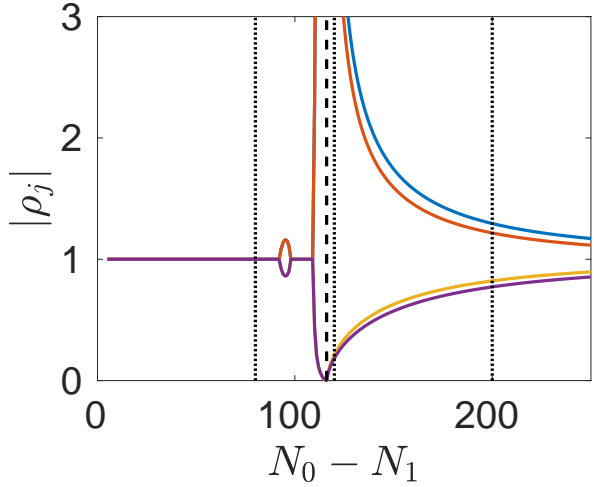}
\caption{\label{fig:CharFloq} (color online) Characteristic Floquet
  multipliers for the orbits of the effective TM model as functions of
  the initial values of $N_0-N_1$. The dashed line marks the critical
  condition $N_0-N_1=116$ for the transition between Josephson and ST
  regimes, and the dotted ones indicate the initial values of
  $N_0-N_1$ for the evolutions shown in Fig.\ \ref{fig:FloIll}.}
\end{figure}

\begin{figure}
\begin{center}
\includegraphics[width=0.9\columnwidth]{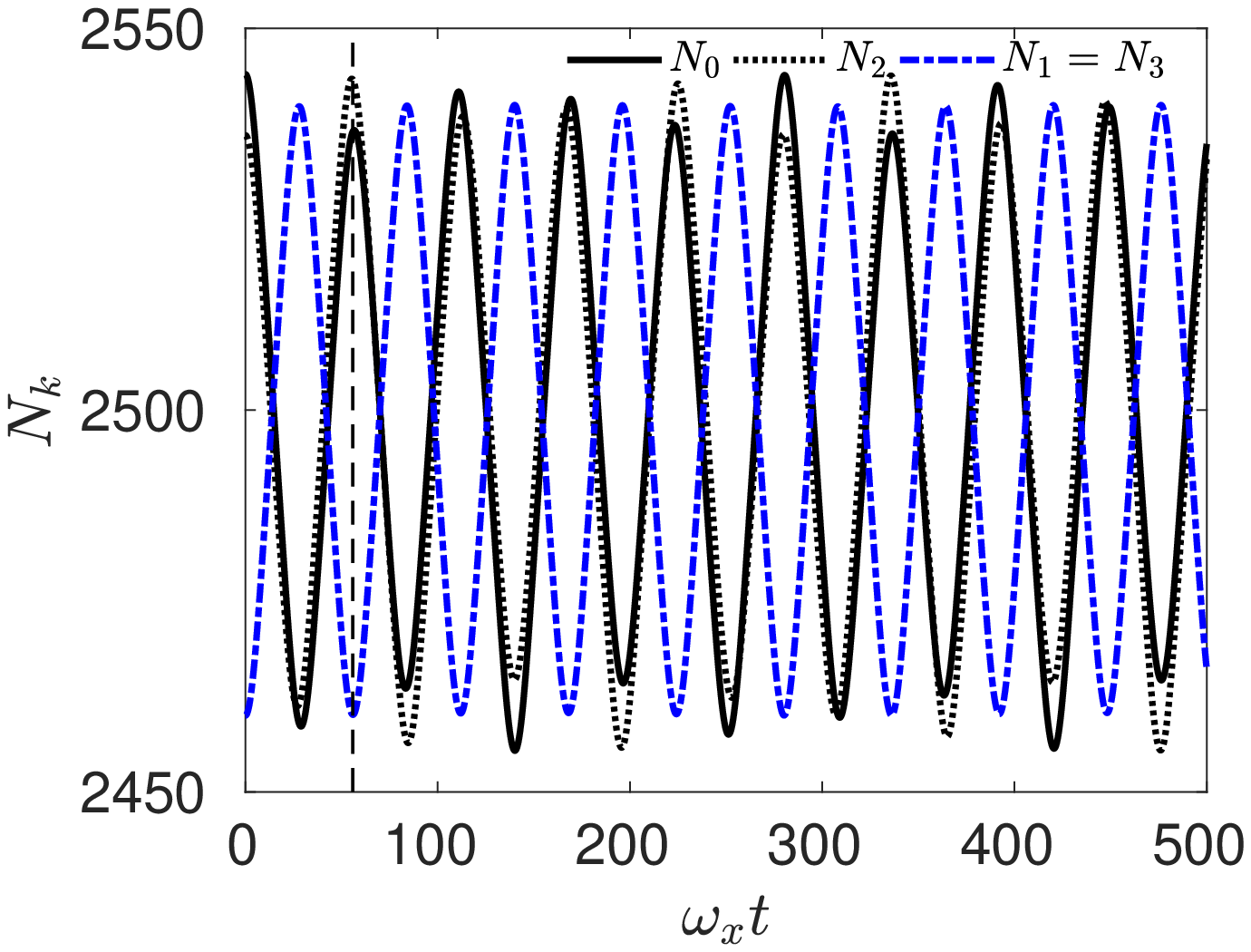} \\
\includegraphics[width=0.9\columnwidth]{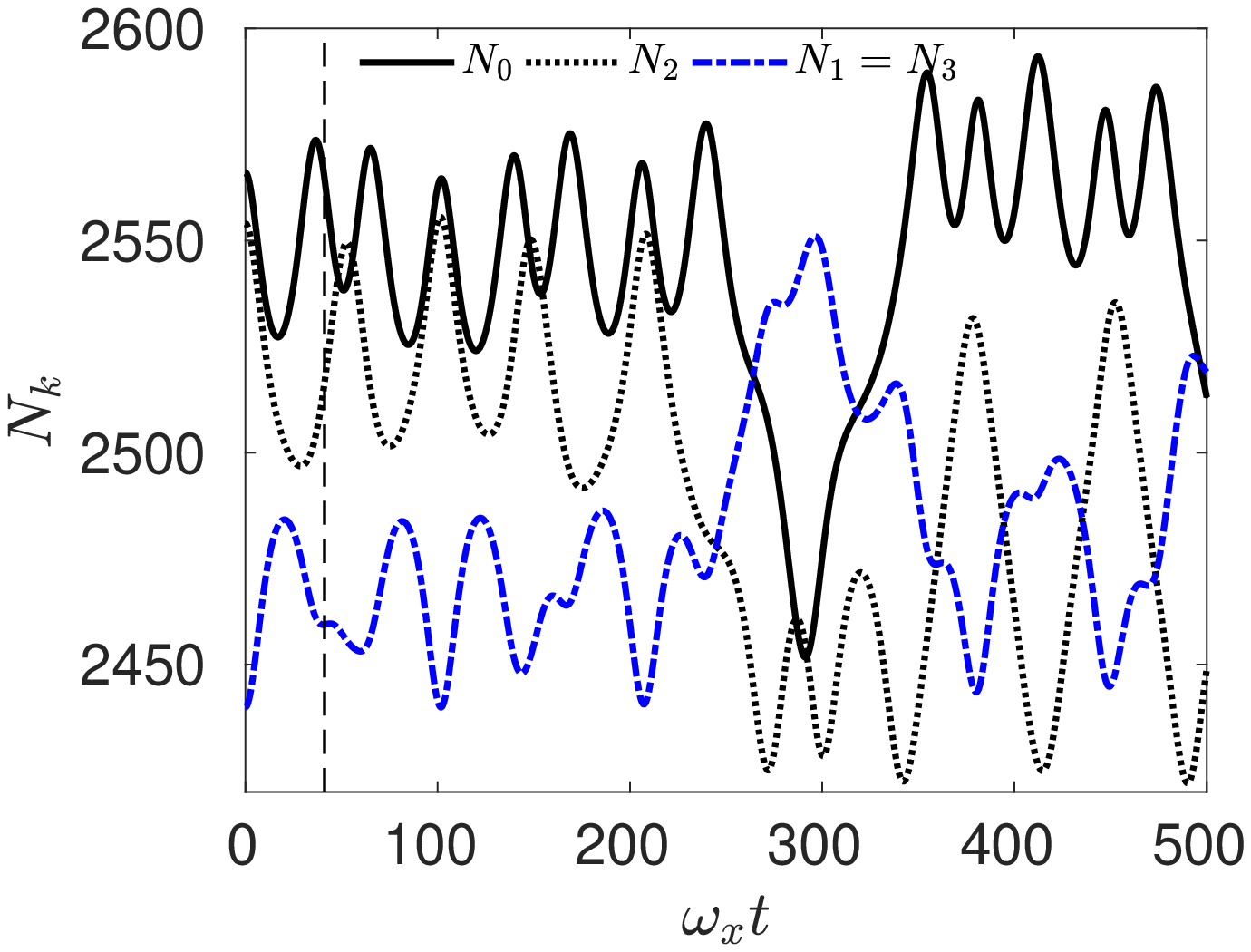} \\
\includegraphics[width=0.9\columnwidth]{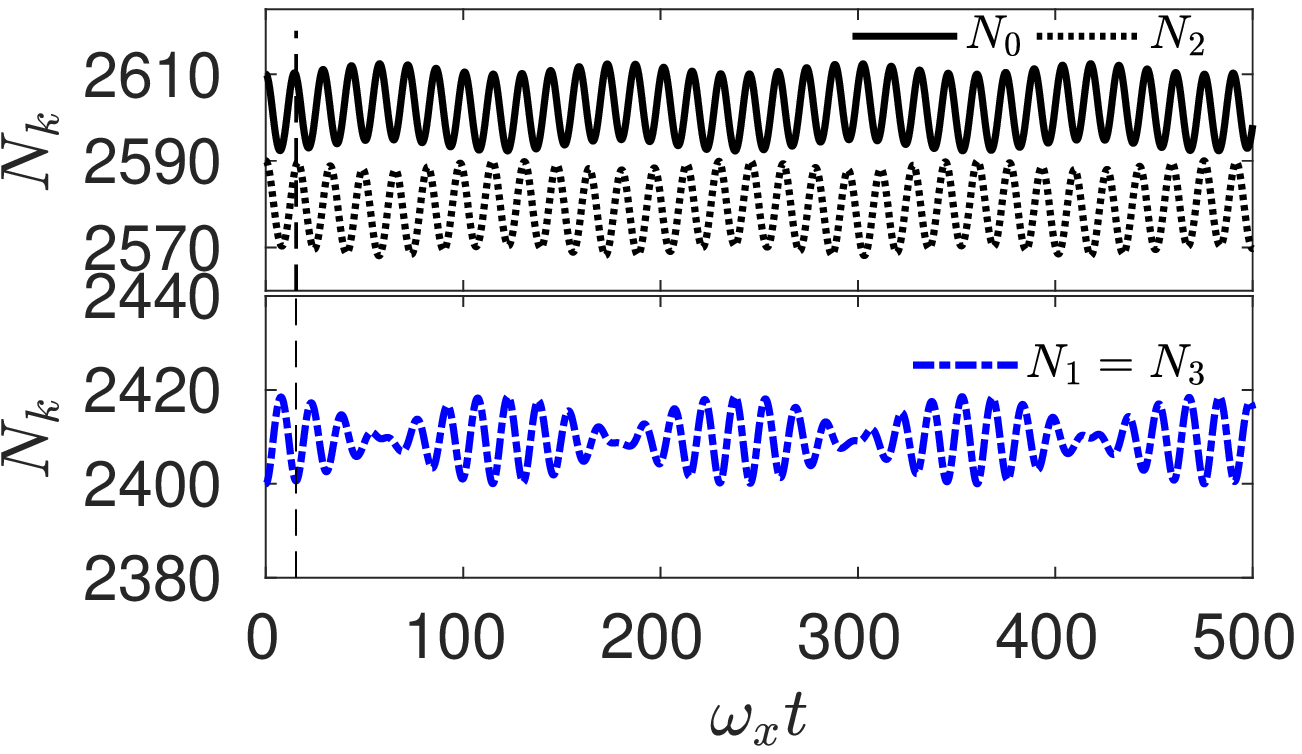} \\
\end{center}
\caption{\label{fig:FloIll}(color online) Time evolution of the
  populations $N_k$ for initial conditions close to the TM symmetric
  case for ${Z}_i=0.016$ (top), $0.024$ (middle) and $0.04$ (bottom),
  and $(n_0-n_2)/(2Z_i)=0.1$. The vertical dashed lines mark the orbit
  periods for the effective TM model at each value of $Z_i$.}
\end{figure}

In the TM symmetric situation of the previous section the system is
governed by the TM Hamiltonian, Eq.\ (\ref{eq:Hred}), and thus the
orbits are periodic. However, for arbitrary non-symmetric initial
conditions the dynamics become non-periodic in general. To investigate
this scenario we perform a linear analysis of the six-dimensional
dynamical system around the TM symmetric case. With this aim, we first
rewrite the time evolution equations in terms of the mean populations
and phases of non-neighboring sites $\bar{n}_{ij}=(n_{i}+n_{j})/2$,
$\bar{\varphi}_{ij}=(\varphi_{i}+\varphi_{j})/2$, and corresponding
differences $\bm\delta$ (see Appendix \ref{sec:appB}). Then,
linearizing the resulting dynamical equations in the differences we
obtain two sets of equations. On the one hand, we recover the TM
equations for ${Z}=2(\bar{n}_{02}-\bar{n}_{13})$ and
$\varphi=\bar{\varphi}_{13}$. And, on the other hand we obtain a
non-autonomous linear system for the differences, which can be cast as
\begin{equation}
  \frac{d \bm{\delta}}{dt} = \mathbb{A}[{Z}(t),\varphi(t)] \bm{\delta},
\label{eq:Adt}
\end{equation}
where the vector $\bm\delta$ is defined as the following differences
\begin{multline}
\bm{\delta} = \frac{1}{2}(n_0-n_2,\varphi_0+\varphi_3-\varphi_1-\varphi_2, \\ n_1-n_3,
\varphi_0+\varphi_1-\varphi_2-\varphi_3).
\end{multline}
The periodicity of ${Z}(t)$ and $\varphi(t)$ gives rise to a Floquet
problem for $\bm{\delta}$ \cite{Floquet}.  Therefore, we shall pursue
the study of the characteristic multipliers $\rho_j$ as functions of
the initial imbalance $Z(0)=Z_i$ and $\varphi(0)=0$. The multipliers
are the eigenvalues of the monodromy matrix $\mathbb{M}$ associated to
Eq.\ (\ref{eq:Adt}) \cite{Floquet} and they contain information on the
evolution of $\bm\delta$ after a period $T$, since
$\bm\delta(T)=\mathbb{M}\cdot \bm\delta(0)$. In particular, each
multiplier $\rho_j$ gives the ratio of change of a linearly
independent solution $\bm{\delta}_j$, i.e.,
$\bm{\delta}_j(T)=\rho_j \bm{\delta}_j(0)$.  Each Floquet multiplier
thus falls into one of the following categories that characterize the
dynamics of the solutions:
\begin{enumerate}
\item If $|\rho_j| < 1$, there is a solution $\bm\delta_j$ asymptotically stable.
\item If $|\rho_j| = 1$, we have a pseudo-periodic solution.  If
  $\rho_j = \pm 1$, then the solution is periodic.
\item If $|\rho_j| > 1$, there is a linearly unstable solution $\bm\delta_j(t)$.
\end{enumerate}
The entire solution is asymptotically stable if all the characteristic
multipliers satisfy $|\rho_j | \le 1$. In Fig.\ \ref{fig:CharFloq} we
show the absolute value of the four Floquet multipliers for the
periodic orbits of Sec.\ \ref{sec:sym}. Given that
$\Tr(\mathbb{A})=0$, and the fact that the matrix $\mathbb{A}$ is
decoupled into blocks of $2\times2$ matrices, the product of the
eigenvalues verify $\rho_1\rho_2=1$ and $\rho_3\rho_4=1$. As it may be
seen, far from the critical imbalance in the Josephson regime the
dynamics is pseudo-periodic, whereas for the ST regime the linear
dynamics is unstable. We also observe a small region,
$0.0185 \lesssim Z \lesssim 0.0195$, where two multipliers exceed the
value 1, indicating that the effect of the instability extends towards
the Josephson regime.  From this analysis one may conclude that near
symmetric initial conditions in the Josephson regime, the dynamics of
the exact model is almost always close to that of the effective TM
model, while for initial conditions around the ST regime the Floquet
analysis predicts linear instability, and thus the non-linearized
dynamics are expected to differ considerably from the effective TM
results. However, for a particular evolution one should also inspect
the involved values of the monodromy matrix elements to understand in
more detail the initial deviation from the TM orbits.  This behavior
is illustrated in Fig.\ \ref{fig:FloIll}, where we plot the numerical
solutions of the EMM dynamics for several initial conditions with a
small $(n_0-n_2)/(2Z)=0.1$, $N_1=N_3$, and zero phases. Furthermore,
it can be easily shown that the choice of zero phases and $N_1=N_3$ as
initial conditions warrants that the dynamics can be described in
terms of four variables only, namely $Z$, $\varphi$ and the
differences $N_0-N_2$ and $\varphi_0-\varphi_2$, instead of the full
six dimensions, as can be expected from the symmetry of such
configuration.

As a consequence of the linear instability, it is not possible to
reliably predict the full dynamics close to $Z_c$ only from the TM
Hamiltonian and a numerical solution of the EMM model must be
performed for each initial condition. This is clearly shown in the
middle panel of Fig.\ \ref{fig:FloIll} with ${Z}_i\gtrsim {Z}_c$,
where the shape of the oscillations of $N_k$ is depicted.
Nonetheless, already for slightly larger values of ${Z}_i$ where all
the $|\rho_j|$ are closer to one, we observe that there is a
characteristic time close to the time period prediction of the TM
model, $T\simeq T^{\mathrm{EMM}}(Z_i)$. In addition to $T$, as seen
clearly in the bottom panel of Fig.\ \ref{fig:FloIll}, there is a
beat-type oscillation with a much longer characteristic time
$T_M\approx 16T$.  This beating can be understood as the composition
of two ST modes for each pair of neighbors with nearby time periods
$T_1$ and $T_2$ given by Eq.\ (\ref{tstbuenar}), leading to the
modulating period $T_M=2T_1T_2/(T_1-T_2)$.

\begin{figure}
  \begin{center}
    \includegraphics[width=\columnwidth]{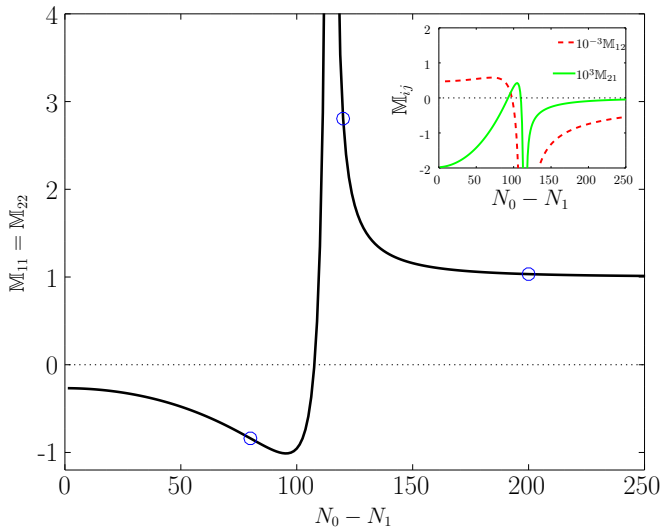}
  \end{center}
  \caption{\label{fig:Monos} (color online) Monodromy matrix elements
    $\mathbb{M}_{ij}$ as functions of the initial imbalance
    $N_0-N_1$. The empty circles mark the values corresponding to
    Fig.\ \ref{fig:FloIll}. }
\end{figure}

The change in the population imbalance after a single period $T$ can
be read off directly from the monodromy matrix. In particular, for the
initial conditions of Fig.\ \ref{fig:FloIll} the element of the matrix
$\mathbb{M}_{11}$ corresponds to the ratio $\delta_1(T)/\delta_1(0)$,
shown in Fig.\ \ref{fig:Monos}. In the Josephson regime for
$Z<0.0215$, $\delta_1(t)$ changes sign and reduces its magnitude after
a period, as can be seen in the top panel of Fig.\ \ref{fig:FloIll},
whereas for $Z_i=0.04$ in the ST regime we observe that $\delta_1(t)$
does not change after a period $T$ (bottom panel of Fig.\ \ref{fig:FloIll}).

\subsection{\label{sec:regN1N3}Regimes in the $N_1=N_3$ symmetric case}
In addition to the investigation of the characteristic times, it is
interesting to wonder whether the system could remain in a Josephson
or in a ST regime in the surrounding region of a given TM symmetric
configuration. Therefore, we have studied the EMM equations of motion
for $n_i(t)$ and $\varphi_i(t)$ for several initial conditions in a
specific plane of the full phase-space containing the symmetric
configuration. In particular, we numerically integrated
Eqs. (\ref{encmode1hn}) and (\ref{encmode2hn}) from $t=0$ to
$t=500\omega_x^{-1}$ with fixed initial phase differences
$\varphi_i=0$ and $N_1=N_3$, while varying $N_0-N_1$ and $N_2-N_3$.
We classify the dynamics for each initial condition in one of the
following three regimes: Josephson (J), mixed (M), and self-trapping
(ST) depending on whether all (J), some (M), or none (ST) of the
populations of neighboring sites cross each other during the time
evolution.

Assuming the double-well condition for a ST regime
(cf. Eq. (\ref{eq:Zct})) for each pair of neighbors forming a
junction, one can obtain a first estimate of the phase-space domains
of the J, M, and ST regimes.  Thus, defining the parameters
\begin{equation}
  A_{ij}=(N_i-N_j)^2 \frac{U_{\text{eff}}}{2K},\;\,\text{and}\;\, B_{ij}=4 (N_i+ N_j) -\frac{8K}{U_{\mathrm{eff}}},
\label{eq:local}
\end{equation}
the populations of neighboring sites $i$ and $j$ should cross each
other when $A_{ij}<B_{ij}$.  Therefore, if this condition is fulfilled
for the four junctions, we expect a J regime; if no junction satisfy
it, we expect a ST regime, and a mixed regime otherwise.  Similar
local two-mode model conditions have been considered in the study of
self-trapping in extended one-dimensional optical lattices
\cite{Anker2005,Wang2006}.

\begin{figure}
\includegraphics[width=1.15\columnwidth,clip=true]{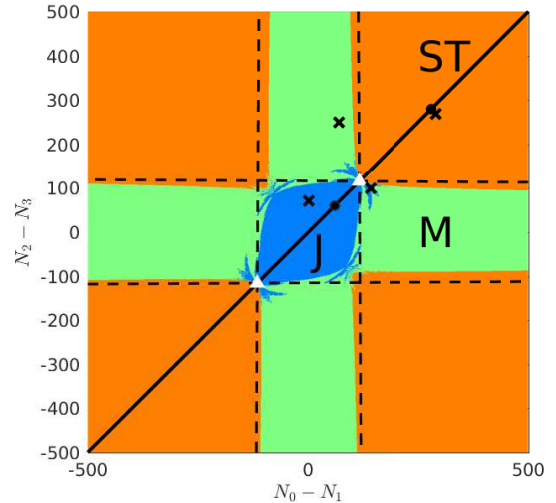}
\caption{\label{fig:PSP} (color online) Phase-space diagram of the
  four-mode model with $\varphi_i=0$ and $N_1=N_3$. The solid line
  marks the symmetric case with $N_0=N_2$ and $N_1=N_3$. The dashed
  lines indicate the local conditions $A_{ij}=B_{ij}$. The triangles
  correspond to the critical condition $N_0-N_1=N_2-N_3=\pm 116$ extracted
  from Eq.~(\ref{eq:Zct}), whereas the circles indicate the symmetric
  initial conditions of Figs.\ \ref{fig:joseph4} and \ref{fig:self4},
  and the crosses the non-symmetric ones in
  Figs. \ref{fig:joseph4M_Mixed}--\ref{fig:pata}. }
\end{figure}

In Fig.\ \ref{fig:PSP} we show a phase-space diagram resulting from
the numerical integration of the EMM dynamics together with the local
predictions implied by Eqs. (\ref{eq:local}). We observe that in the
neighborhood of a symmetric configuration not immediate to the
critical point, the character of the regime does not change,
\text{i.e.} a ST regime remains as ST, and so does a J regime. Only in
the close vicinity of the critical point in the symmetric
configuration, we numerically observe that the dynamics possesses
qualitative details not contained in the local analysis of Eq.\
(\ref{eq:local}). This result was expected from the Floquet analysis
of the previous section, where the absolute value of some multipliers
greatly exceeded 1 around $Z_c$. The detailed study of such
phase-space regions deserve a separate and careful numerical analysis
which lies beyond the scope of the present work.

\begin{figure}
\includegraphics[width=0.9\columnwidth,clip=true]{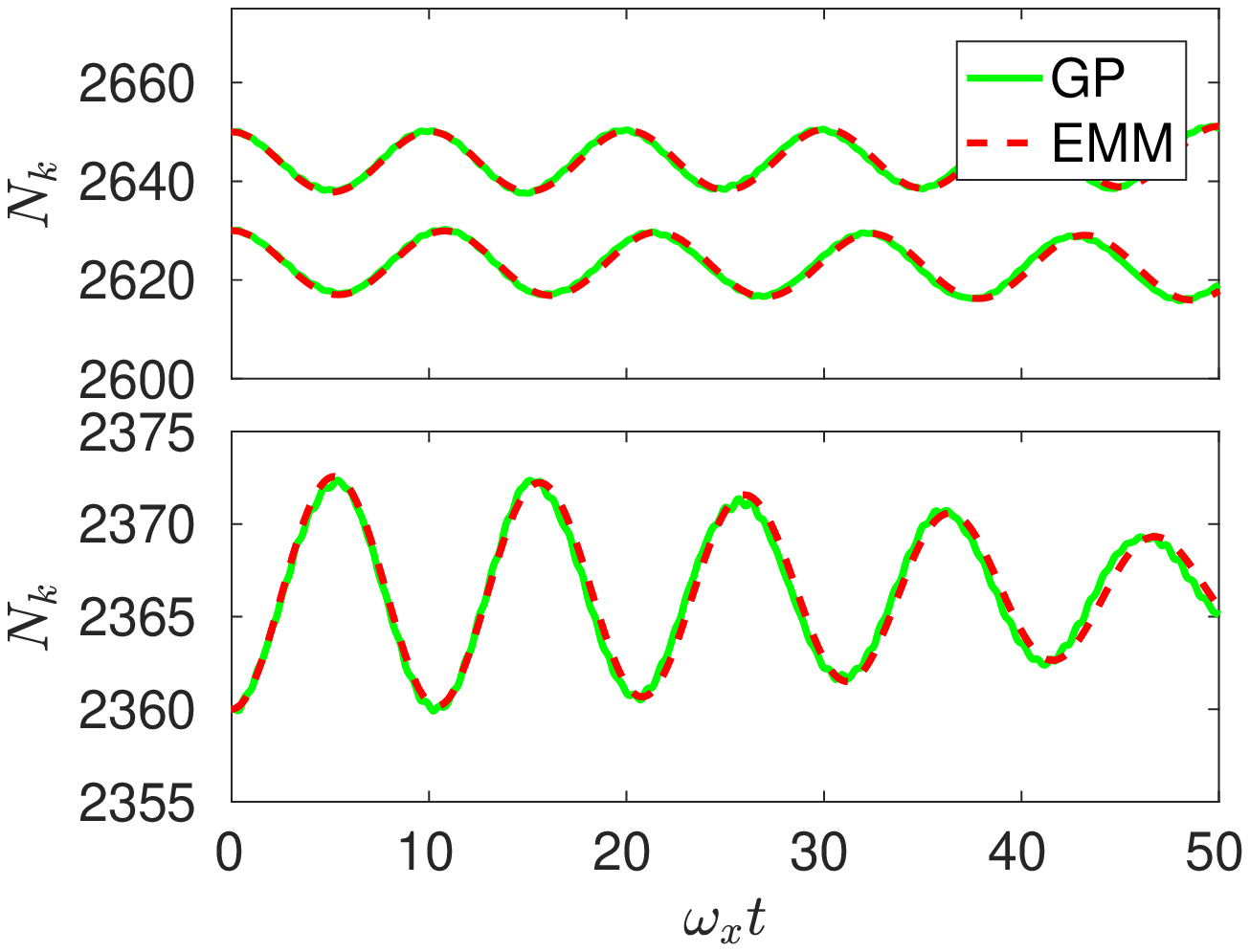}
\includegraphics[width=0.9\columnwidth,clip=true]{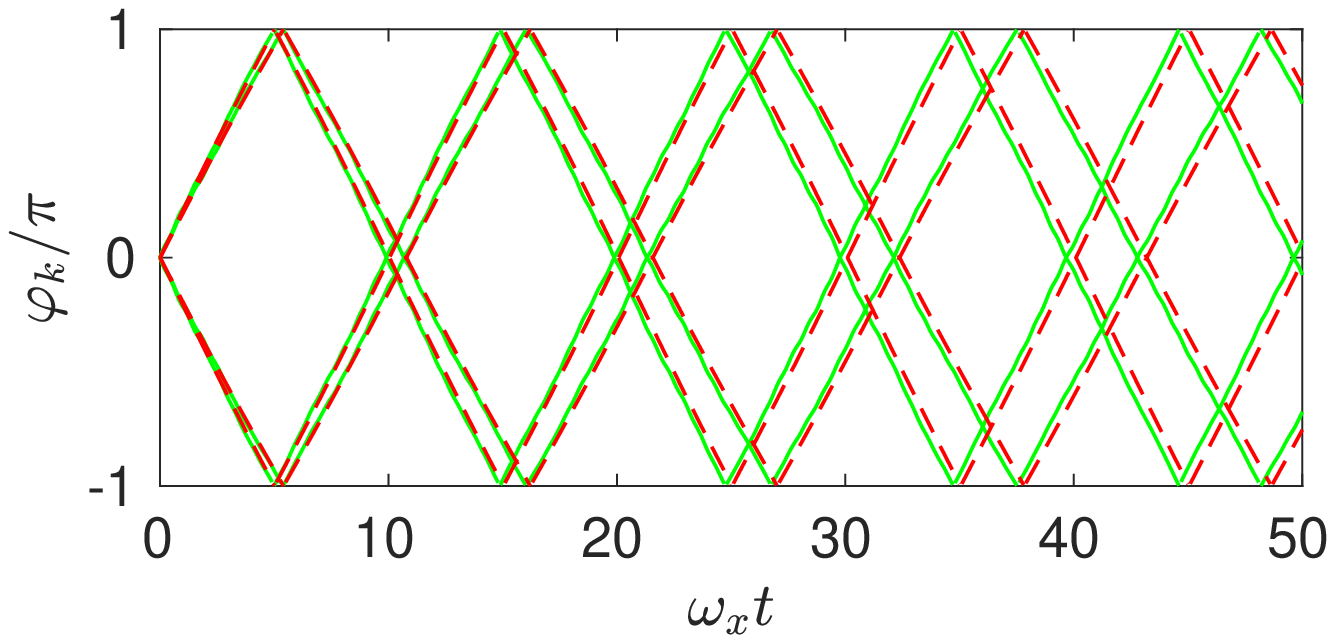}
\caption{\label{fig:joseph4M_Mixed} (color online) Dynamics arising
  from the GP simulation and the EMM model, for initial values
  $N_0=2650, N_1=2360, N_2=2630$, and $N_3=2360$, with
  $V_b/(\hbar\omega_x)=25$. Top panel: populations $N_k$, bottom
  panel: phase differences $\varphi_k$.}
\end{figure}

\begin{figure}
  \includegraphics[width=0.9\columnwidth,clip=true]{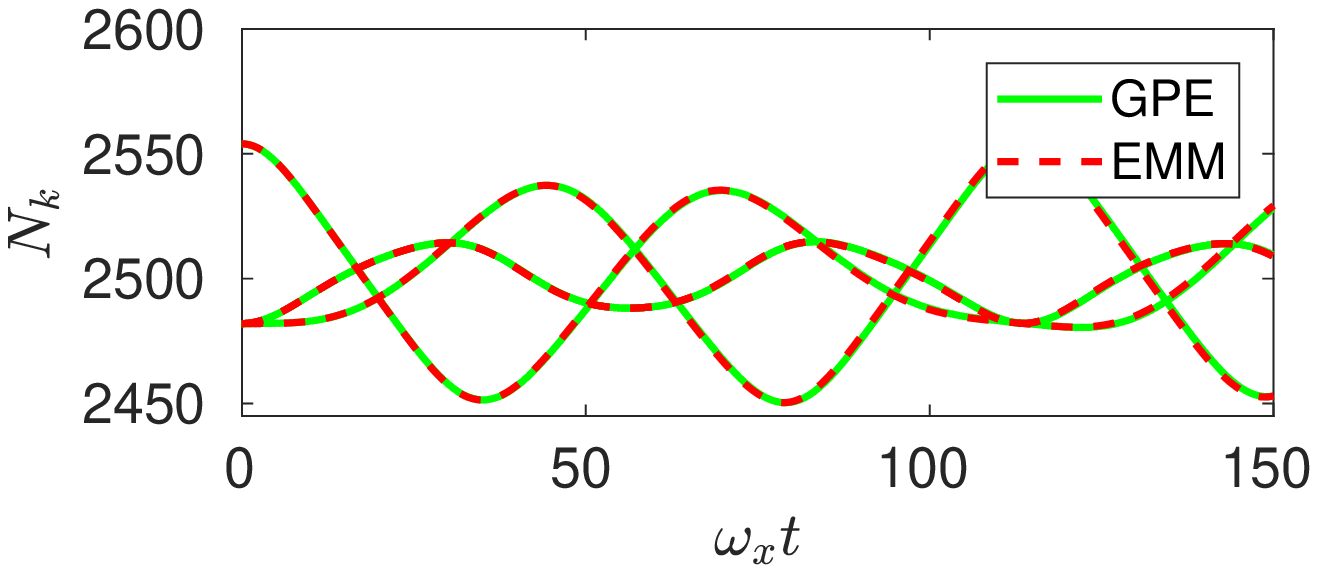}
  \includegraphics[width=0.9\columnwidth,clip=true]{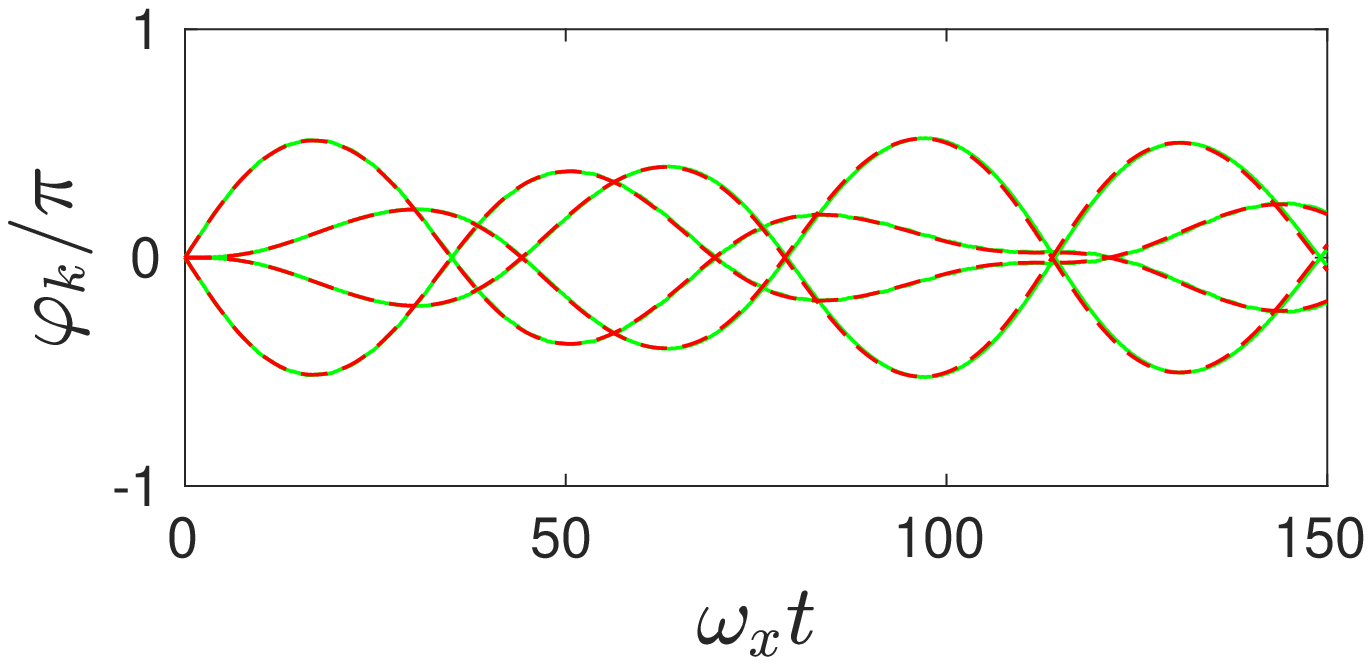}
\caption{\label{fig:dentro}(color online) Same as Fig.\
  \ref{fig:joseph4M_Mixed} for $N_0=2482$, $N_1=2482$, $N_2=2554$, and $N_3=2482$. }
\end{figure}

To verify our findings within the framework of the full GP equations,
we have numerically solved several non-symmetric initial conditions.
First, we slightly move from the symmetric point in the ST regime in
Fig.\ \ref{fig:self4}, and choose $ N_0 = 2650$, $N_1=2360$,
$ N_2 = 2630$, and $N_3=2360$. We show in Fig.\
\ref{fig:joseph4M_Mixed} the corresponding time evolution.  We note
that the characteristic times in the surrounding region of the
symmetric case smoothly depart from the orbit periods of the two-mode
models. In particular, it can be seen that in
Fig.~\ref{fig:joseph4M_Mixed} the oscillation times keep around
$10 \omega_x^{-1}$, very close to that of Fig.\ \ref{fig:self4},
although this simulation does not correspond to a closed
orbit. Similarly in Fig.\ \ref{fig:dentro} we compare the time
evolutions in the Josephson regime within the GP and EMM approaches,
finding an excellent agreement.

\begin{figure}
\includegraphics[width=0.9\columnwidth,clip=true]{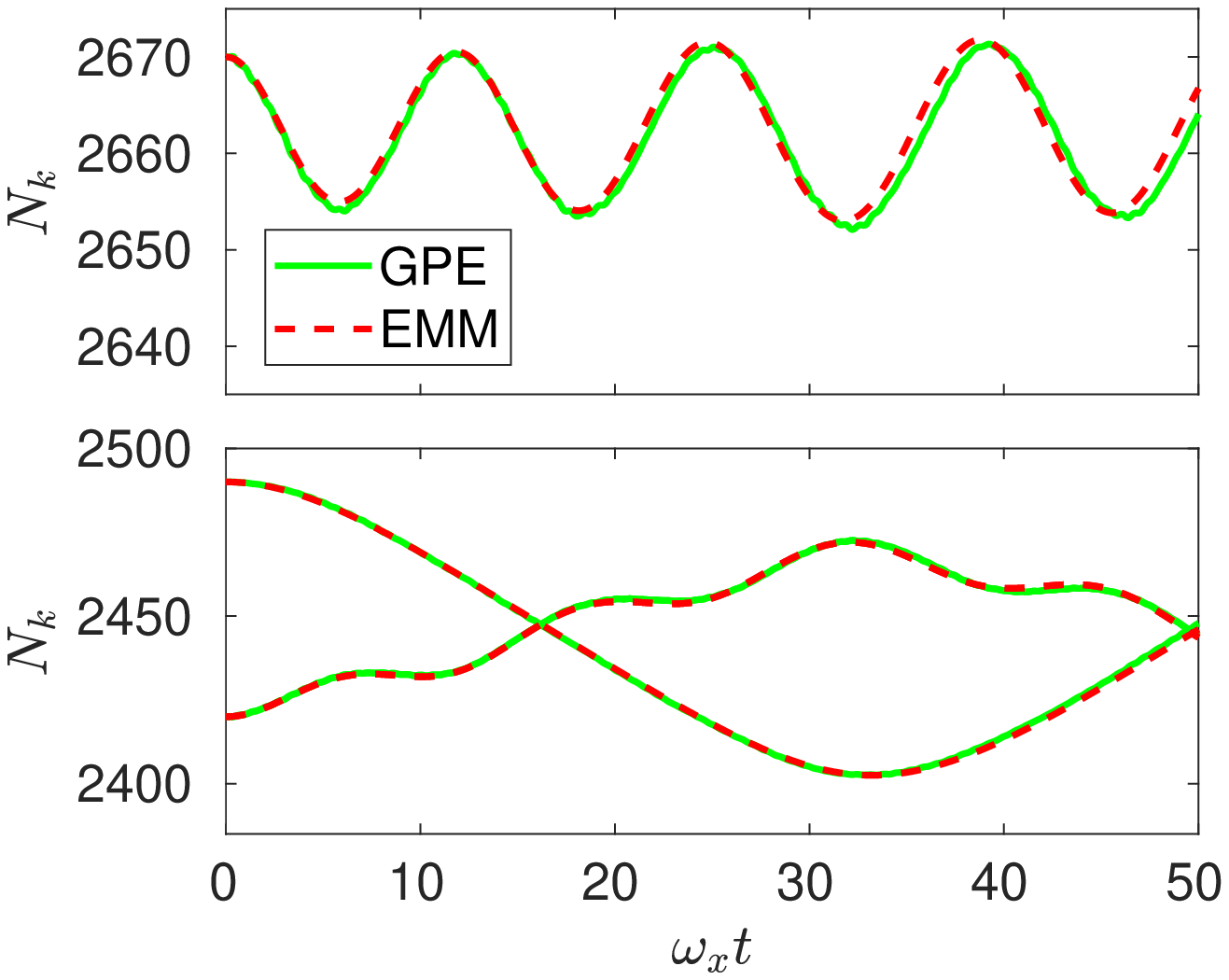}
\includegraphics[width=0.9\columnwidth,clip=true]{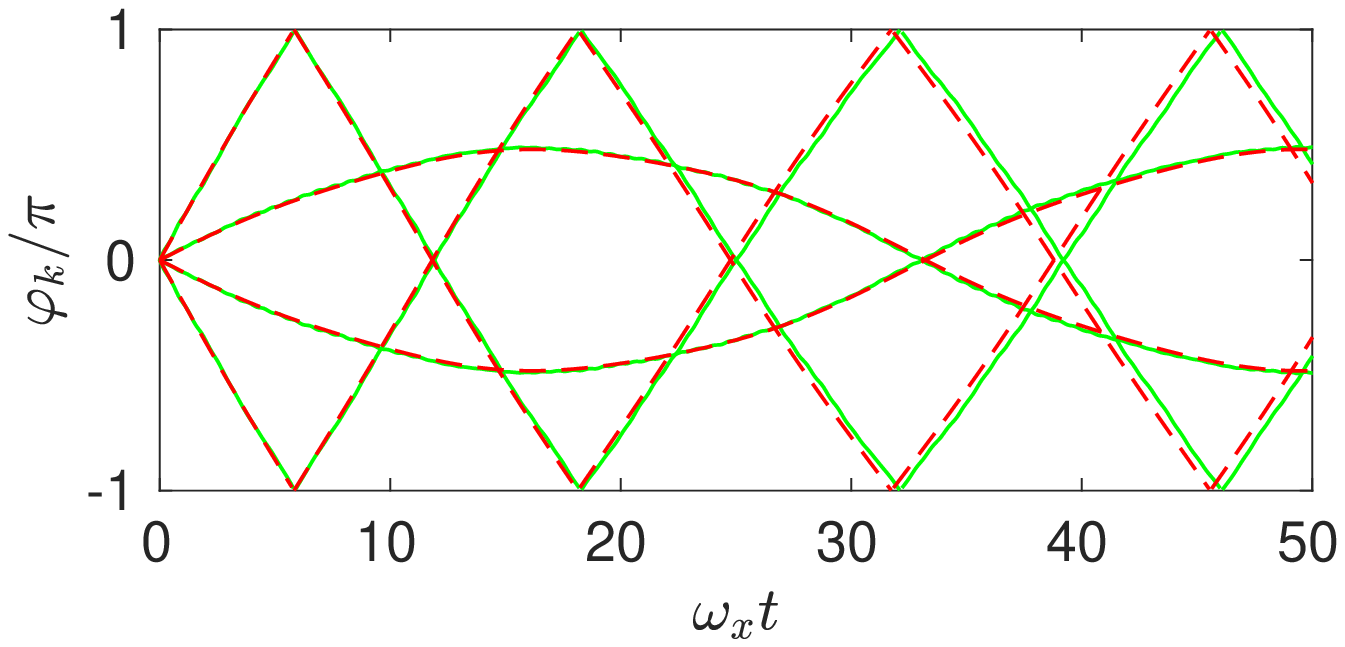}

\caption{\label{fig:joseph4M_Mixed3}(color online) Same as Fig.\
  \ref{fig:joseph4M_Mixed} for $N_0=2490, N_1=2420, N_2=2670$, and
  $N_3=2420$. }
\end{figure}

We present in Fig.\ \ref{fig:joseph4M_Mixed3} another example of a
non-symmetric initial condition within the GP equations, where we
expect that this time it will be in a mixed regime. We take the
initial populations $N_0=2490, N_1=2420, N_2=2670$ and $N_3=2420$.
Note that in Fig.\ \ref{fig:joseph4M_Mixed3} some of the initial
populations differences lie below the critical threshold obtained for
the symmetric case, and these conditions turned out to be useful for
predicting that this state would be in a mixed regime also in the GP
framework.

\begin{figure}
  \includegraphics[width=0.9\columnwidth,clip=true]{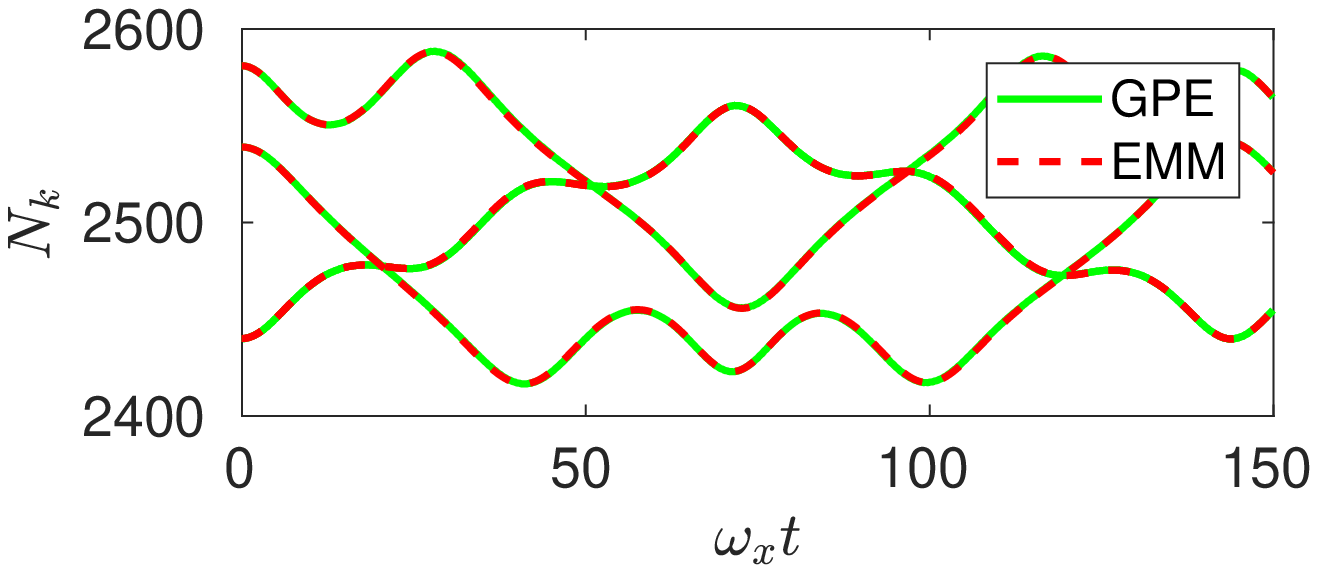}
  \includegraphics[width=0.9\columnwidth,clip=true]{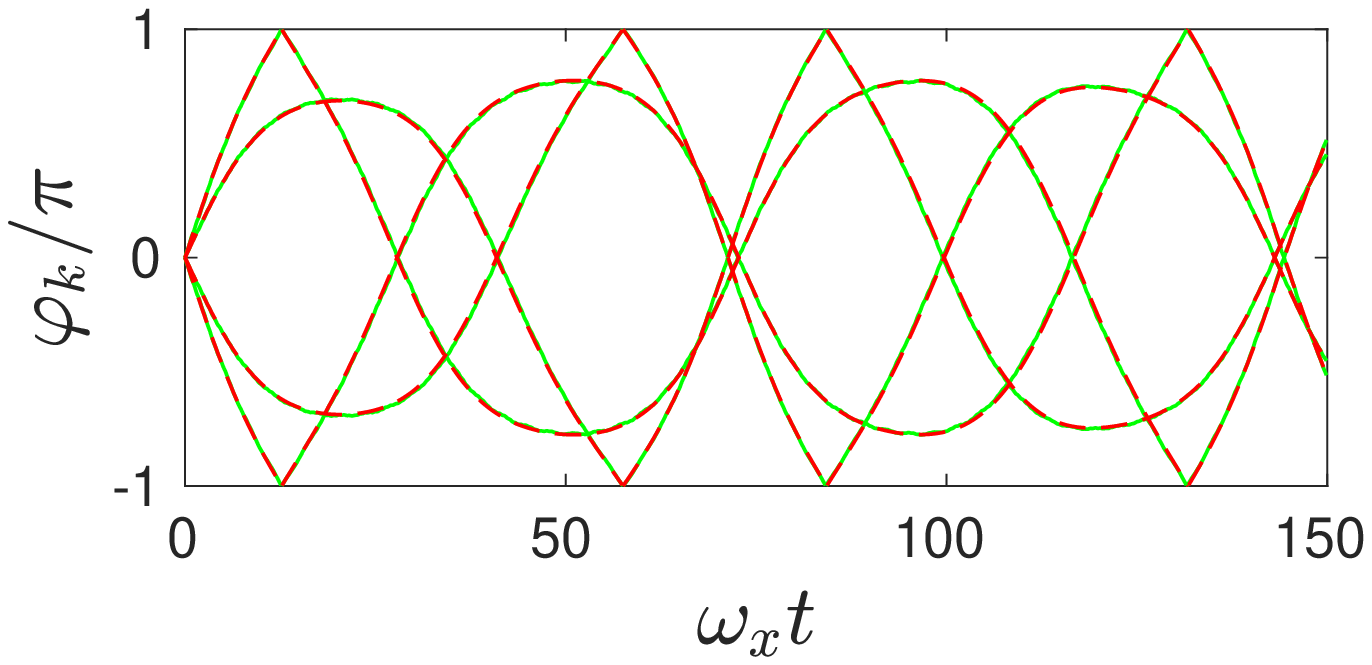}
\caption{\label{fig:pata}(color online) Same as Fig.\
  \ref{fig:joseph4M_Mixed} for $N_0=2581$, $N_1=2440$, $N_2=2539$, and $N_3=2440$. }
\end{figure}

Finally, in Fig.\ \ref{fig:pata} we show the GP evolution for a
particular initial condition which, according to the simplified local
model, should have been in the mixed regime but, as predicted within
the EMM model, it is in a Josephson regime.

To conclude, from the previous analysis it can be seen that the study
of symmetric configurations is a convenient starting point for
exploring the full six dimensional system ($n_i$,$\varphi_i$) arising
from the four-mode model. We have shown that the adoption of TM tools
allows us to predict different regimes and characteristic times for
the dynamics of four-site condensates. These results may be easily
extended to condensates with a larger even number of sites.

\section{\label{sum} Summary and concluding remarks}

We have studied the dynamics of three-dimensional four-well
Bose-Einstein condensates using a multimode model with an effective
interaction parameter, and compared it to the Gross-Pitaevskii
solutions. In order to predict orbits in four-well systems and to
establish the characteristic time scales of the dynamics, we have
first studied a highly symmetric configuration that admits a two-mode
Hamiltonian description.  This allowed us to apply previous two-mode
results to this reduced-space case. Moreover, the location of the
critical point marking the transition from Josephson to self-trapping
regimes in the symmetric case has shown to be useful for defining in
the extended phase space, zones with different dynamical regimes. For
general initial conditions close to the two-mode symmetric ones, we
performed a Floquet analysis that revealed that the linearized
dynamics is unstable around the critical imbalance points.  The linear
instability in the Floquet problem explains the coexistence of
different regimes in the neighborhood of the critical imbalance in the
extended phase space.

We have characterized the possible dynamical regimes according to the
population behavior between neighboring sites. Hence, we have
performed a partition of an extended region of the phase space into
self-trapped, Josephson, and mixed regimes. We have confirmed these
findings by an extensive numerical study of non-symmetric initial
conditions within the multimode model, together with a set of time
evolutions within the three-dimensional Gross-Pitaevskii framework.

We must emphasize that the accuracy of the predictions of the
multimode model depends on the proper determination of its parameters.
On the one hand, we have obtained very good agreements due to the use
of the effective on-site interaction energy parameter
$U_{\text{eff}}$. Such a parameter amounts to a reduction of about a
factor 0.72 with respect to the bare $U$ of Eq.\ (\ref{U0}). On the
other hand, we have also shown that $U$ strongly depends on the
localization of the underlying Wannier-like functions. Such a
localization has to be maximized in order to obtain an accurate
model. This can be achieved by minimizing its spatial dispersion with
respect to a parameter that defines the global phases of the
stationary states with $n=\pm 1$ winding numbers.  This procedure
should also be applied to systems with a larger number of sites, where
more stationary states with non zero velocity circulations exist, and
therefore another variational parameter has to be added for each new
absolute value of the winding number.

To conclude we would like to remark that the present study of
four-well systems could pave the way for an eventual generalization to
condensates with a higher number of sites.

\begin{acknowledgments}
 This work was supported by CONICET and Universidad de Buenos
 Aires through grants PIP 11220150100442CO and UBACyT
 20020150100157BA,
 respectively.
 \end{acknowledgments}

\appendix

\section{Multimode Parameters \label{sec:parameters}} 

The parameters of the MM mode are defined as

\begin{equation}
J= -\int d^3{\bf r}\,\, w_0 ({\bf r}) \left[
-\frac{ \hbar^2 }{2 m}{\bf \nabla}^2  +
V_{\rm{trap}}({\bf r})\right]  w_1({\bf r})
\label{jota0}
\end{equation}
\begin{equation}
U= g \int d^3{\bf r}\,\,  w_0({\bf r})^4
\label{U0}
\end{equation}
\begin{equation}
F= -  g N \int d^3{\bf r}\,\,  w_0^3({\bf r})
 w_1 ({\bf r})
\label{jotap0}
\end{equation}
The parameter $F$ were first defined in Ref. \cite{anan06} and later
analyzed in Ref. \cite{jia08}.

Together with the calculation of these parameters through the
preceding definitions, we have applied also the alternative method
outlined in Ref. \cite{cat11}, finding an agreement between both
procedures with a precision higher than 99\%.  For the MM model
applied to the four-well system, the hopping parameter can be written
as
\begin{equation}
\Delta E = E_2 - E_0 =  4J+ 2 F = 2  K   \,,
\label{tstbuena}
\end{equation}
where $E_2$ is the energy of the stationary state with winding number
$n=2$ and $E_0$ is the energy of the ground state.  It is interesting
to note that one can also assure that we have used high enough
barriers, provided the parameters calculated through the definitions
(\ref{jota0})--(\ref{jotap0}) are in agreement with the corresponding
results arising from the above alternative method.  For example, in
our numerical calculation $ V_b= 25 $ $ \hbar \omega_x $ and we get
$ E_2 - E_0 = 1.5251 \times 10^{-3}$ $ \hbar \omega_x $ , whereas by
definition we obtained $ 2 K=1.525 \times 10^{-3} $
$ \hbar \omega_x $.  On the other hand, if using $V_b=15$
$ \hbar \omega_x $, we get $ E_2 - E_0 = 8.78 \times 10^{-2} $
$ \hbar \omega_x $, being by definition $ 2 K=8.85 \times 10^{-2} $
$ \hbar \omega_x $, which clearly reflects a less accurate model.

\section{\label{sec:appB}Linearized dynamics around periodic orbits}

Starting from the dynamical equations
(\ref{encmode1hn})-(\ref{encmode2hn}) we make a change of variables to the mean values
\begin{align}
&\bar{n}_{02}=(n_0+n_2)/2,\quad& \bar{n}_{13}=(n_1+n_3)/2,\nonumber \\ 
&\bar{\varphi}_{02}=(\varphi_0+\varphi_2)/2,\quad&
\bar{\varphi}_{13}=(\varphi_1+\varphi_3)/2,
\end{align}
and the differences
\begin{align}
&\delta_1 =(n_0-n_2)/2, &\delta_3 &=(n_1-n_3)/2, \nonumber \\
&\delta_2 =(\varphi_0+\varphi_3-\varphi_2-\varphi_1)/2,
&\delta_4 &=(\varphi_0+\varphi_1-\varphi_2-\varphi_3)/2.
\end{align}
Linearizing the resulting equations on the differences $\delta_i$ and
taking into account that $\bar{n}_{02}+\bar{n}_{13}=N/2$ and
$\bar{\varphi}_{02}=-\bar{\varphi}_{13}$, we obtain two sets of
equations. On the one hand, the mean values are decoupled from the
differences $\bm\delta$ and thus there are only two independent
equations, namely,
\begin{align}
  \hbar\dot{\bar{n}}_{02}=&\frac{K}{2}  \sin\bar{\varphi}_{02} \sqrt{1 - Z^2}, \\
\hbar\dot{\bar{\varphi}}_{02} =&-\frac{U_{\mathrm{eff}}N}{2} Z - 2 K Z\frac{\cos\bar{\varphi}_{02}}{\sqrt{1-Z^2}}.
\label{eq:A2:Z}
\end{align}
On the other hand, for $\delta_i$ we obtain
\begin{widetext}
  \begin{align}
    \hbar\, \dot\delta_1 &=   -\sqrt{1-Z^2} \left(F + \frac{K}{1+Z}\right)\sin\varphi\, \delta_1 + \frac{1}{4}\sqrt{1-Z^2} K \cos\varphi\,\delta_2 \nonumber \\
    \hbar\, \dot\delta_2 &= \left[-2 \left(U_{\text{eff}} -2\alpha Z U\right)N + 4 \frac{1-Z}{\sqrt{1-Z^2}}\left(2 F - \frac{K}{1+Z}\right)\cos\varphi\right]\,\delta_1 + \sqrt{1-Z^2}\left(F+\frac{K}{1+Z}\right)\sin\varphi\,\delta_2 
    \label{eq:dn1}
  \end{align}
  \begin{align}
    \hbar\,\dot\delta_3 &=   \sqrt{1-Z^2} \left(F + \frac{K}{1-Z}\right)\sin\varphi\, \delta_3 + \frac{1}{4}\sqrt{1-Z^2} K \cos\varphi\,\delta_4 \nonumber \\
    \hbar\,\dot\delta_4 &= \left[-2 \left(U_{\text{eff}} +2\alpha Z U\right)N + 4 \frac{1+Z}{\sqrt{1-Z^2}}\left(2 F - \frac{K}{1-Z}\right)\cos\varphi\right]\,\delta_3 - \sqrt{1-Z^2}\left(F+\frac{K}{1-Z}\right)\sin\varphi\,\delta_4 
    \label{eq:dn2}
  \end{align}
\end{widetext}
where $Z=Z(t)=2(\bar{n}_{02}-\bar{n}_{13})$ and
$\varphi(t)=-\bar{\varphi}_{02}=-\bar{\varphi}_{02}(t)$ are the
solutions of Eq.~(\ref{eq:A2:Z}) for a periodic orbit of the TM
model. The equations for $\bm\delta$ can thus be cast as the matrix
equation
\begin{equation}
\frac{d\bm\delta}{dt}=\mathbb{A}[Z(t),\varphi(t)]\bm\delta.
\label{eq:A2:deltaeq}
\end{equation}
where the matrix $\mathbb{A}$ is formed by two uncoupled blocks of
$2\times2$ (cf. Eqs.\ (\ref{eq:dn1})--(\ref{eq:dn2})) for the
$(\delta_1,\delta_2)$ and $(\delta_3,\delta_4)$ differences,
respectively. Given the periodicity of $Z(t)$ and $\varphi(t)$ the
dynamical system defines a Floquet problem \cite{Floquet}. To analyze
the stability of the system, we first construct the associated
fundamental monodromy matrix. This can be constructed from the
solution of Eq.\ (\ref{eq:A2:deltaeq}) evaluated at $t=T$, obtained
from the solutions for the set of 4 initial conditions
$\bm\delta_i(0)=\hat{e}_i$ with $\hat{e}_i$ the canonical vectors. The
four solutions $\bm\delta_i(t)$ are then used as columns to form the
monodromy matrix $\mathbb{M}$,
\begin{equation}
\mathbb{M} = \left( \bm\delta_1(T)|\bm\delta_2(T)|\bm\delta_3(T)|\bm\delta_4(T)\right).
\end{equation}
Finally, the diagonalization of $\mathbb{M}$ provides the four Floquet characteristic
multipliers.


\begin{thebibliography}{99}
%
\bibitem{amico1} L. Amico, A. Osterloh, and F. Cataliotti, 
Phys. Rev. Lett. \textbf{95}, 063201 (2005).
%
\bibitem{hen09}
K. Henderson, C. Ryu, C. MacCormick, and M. G. Boshier,
New J. Phys. \textbf{11}, 043030 (2009).
%
\bibitem{jen16}
F. Jendrzejewski,
S. Eckel, T. G. Tiecke, G. Juzeliunas, G. K. Campbell, Liang Jiang, and A. V. Gorshkov, Phys. Rev. A
 \textbf{94}, 063422 (2016).
%
\bibitem{zurek}
J. Dziarmaga, M. Tylutki, and W. H. Zurek, Phys. Rev. B
 \textbf{84}, 094528 (2011); Phys. Rev. B \textbf{86}, 144521 (2012).
%
\bibitem{amico} L. Amico, D. Aghamalyan,
F. Auksztol, H. Crepaz, R. Dumke, L. C. Kwek, Sci. Rep. 
\textbf{4}, 4298 (2014).
%
\bibitem{trespozos2011}  T. F. Viscondi and K. Furuya, J. Phys. A: Math. Theor. \textbf{44},
 175301 (2011).
%
\bibitem{cuatropozos06} S. De Liberato and C. J. Foot,
 Phys. Rev. A \textbf{73}, 035602 (2006).
%
\bibitem{jezek13b} D. M. Jezek  and H. M. Cataldo, Phys. Rev. A
 \textbf{88}, 013636 (2013).
%
\bibitem{smerzi97} A. Smerzi, S. Fantoni, S. Giovanazzi, and S. R. Shenoy, 
Phys. Rev. Lett. \textbf{79}, 4950 (1997).
%
\bibitem{ragh99}
S. Raghavan, A. Smerzi, S. Fantoni, and S. R. Shenoy, Phys. Rev. A
 \textbf{59}, 620 (1999).
\bibitem{anan06}
D. Ananikian and T. Bergeman, Phys. Rev. A
 \textbf{73}, 013604 (2006).
%
\bibitem{jia08}
Xin Yan Jia, Wei Dong Li, and J. Q. Liang, Phys. Rev. A
 \textbf{78}, 023613 (2008).
%
\bibitem{albiez05} M. Albiez, R. Gati, J. F\"olling, S. Hunsmann, M. Cristiani, 
and M. K. Oberthaler, Phys. Rev. Lett. \textbf{95}, 010402 (2005).
%
\bibitem{mele11} M. Mel\' e-Messeguer, B. Juli\'a-D\'{\i}az, M. Guilleumas, 
A. Polls and A. Sanpera, New J. Phys. \textbf{13}, 033012  (2011).
%
\bibitem{abad11}
M. Abad, M. Guilleumas, R. Mayol, M. Pi, and D. M. Jezek,  
Europhys. Lett. \textbf{94}, 10004 (2011).
%
\bibitem{doublewell} T. Mayteevarunyoo, B. A. Malomed, and G. Dong, 
Phys. Rev. A \textbf{78}, 053601 (2008);
B. Xiong, J. Gong, H. Pu, W. Bao, and B. Li, Phys. Rev. A \textbf{79},
 013626 (2009);
Qi Zhou, J. V. Porto, and S. Das Sarma,  Phys. Rev. A
 {\textbf 84}, 031607 (2011); 
B. Cui, L.C. Wang, and X. X. Yi, Phys. Rev. A
 \textbf{82}, 062105 (2010);
 M. Abad, M. Guilleumas, R. Mayol, M. Pi, and D. M. Jezek, Phys. Rev. A
  \textbf{84}, 035601 (2011). 
%
\bibitem{optlat} C. E. Creffield, Phys. Rev. A \textbf{75}, 031607(R) (2007). 
Ju-Kui Xue, Ai-Xia Zhang, and Jie Liu,  Phys. Rev. A \textbf{77}, 013602 (2008).
 T. J. Alexander, E. A. Ostrovskaya, and Y. S. Kivshar, 
Phys. Rev. Lett. \textbf{96}, 040401 (2006). Bin Liu, Li-Bin Fu, Shi-Ping Yang, and Jie Liu,
Phys. Rev. A \textbf{75}, 033601 (2007).
%
\bibitem{Anker2005} Th Anker, M. Albiez, R. Gati, S. Hunsmann,
 B. Eiermann, A. Trombettoni, and M. K. Oberthaler,
 Phys. Rev. Lett. 94, 020403 (2005).

\bibitem{Wang2006} Bingbing Wang, Panming Fu, Jie Liu, and Biao Wu,
 Phys. Rev. A 74, 063610 (2006).

\bibitem{stlastoplat}
A. R. Kolovsky, Phys. Rev. A \textbf{82}, 011601(R) (2010), S. K. Adhikari,
 J. Phys. B: At. Mol. Opt. Phys. \textbf{44}, 075301 (2011).
%
\bibitem{jezek13a} D. M. Jezek, P. Capuzzi,  and H. M. Cataldo, Phys. Rev. A
 \textbf{87}, 053625 (2013).
%
%
\bibitem{nosEPJD} Mauro  Nigro, Pablo  Capuzzi, Horacio M. Cataldo, and Dora M.
  Jezek, Eur. Phys. J. D \textbf{71}, 297 (2017).
%
\bibitem{Floquet} C. Chicone, Ordinary Differential Equations with
  Applications, 2nd ed. (Springer, New York, 2006).
%
\bibitem{cat11}
H. M. Cataldo and D. M. Jezek, Phys. Rev. A \textbf{84},  013602 (2011).
%
\bibitem{mar12} N. Marzari, A. A. Mostofi, J. R. Yates,  I. Souza,  and D. Vanderbilt,
Rev. Mod. Phys. \textbf{84}, 1419 (2012).
%
\bibitem{gros61}
E. P. Gross, Nuovo Cimento \textbf{20}, 454 (1961);
L. P. Pitaevskii, Zh. Eksp. Teor. Fiz.  {\bf 40}, 646 (1961) 
[Sov. Phys. JETP \textbf{13}, 451 (1961)].
%
\bibitem{je11}
D. M. Jezek and H. M. Cataldo, Phys. Rev. A \textbf{83}, 013629 (2011).
%


\end{thebibliography}
\end{document}